\documentclass[useAMS,usenatbib]{mnras}
\usepackage{mathptmx}
\usepackage{color}
\usepackage{graphics,graphicx,amssymb,amsmath,natbib}
\usepackage{deluxetable}
\usepackage{float}
\usepackage{color}
\usepackage{lscape,graphicx}
\usepackage{amsmath}
\usepackage{amssymb}
\usepackage{multirow}
\usepackage{afterpage}
\usepackage{subfig}
\usepackage{lscape}
\usepackage[flushleft]{threeparttable}

\def\cha{{\em Chandra}}
\def\xmm{{\em XMM-Newton}}
\def\rosat{{\em ROSAT}}
\def\hst{{\em HST}}

\title[Cluster/QSO pair]{Probing the dynamical state, baryon content, and multiphase nature of galaxy clusters with bright background QSOs}
\author
[Chong Ge et al.]{Chong Ge$^{1,2,3,4}$, Q. Daniel Wang$^{5}$\thanks{E-mail: wqd@astro.umass.edu}, Joseph N. Burchett$^5$, Todd M. Tripp$^5$, Ming Sun$^{4}$, 
\newauthor
Zhiyuan Li$^{3,6,7}$, Qiusheng Gu$^{3,6,7}$ and Li Ji$^{1,2}$\\
\\
$^{1}$Purple Mountain Observatory, Chinese Academy of Sciences, Nanjing 210008, China\\
$^{2}$Key Laboratory of Dark Matter and Space Astronomy, PMO, CAS, Nanjing 210008, China\\
$^{3}$Key Laboratory of Modern Astronomy and Astrophysics (Nanjing University), Ministry of Education, Nanjing 210093, China\\
$^{4}$Department of Physics and Astronomy, University of Alabama in Huntsville, Huntsville, AL 35899, USA\\
$^{5}$Department of Astronomy, University of Massachusetts, Amherst, MA 01003, USA\\
$^{6}$School of Astronomy and Space Science, Nanjing University, Nanjing 210093, China\\
$^{7}$Collaborative Innovation Center of Modern Astronomy and Space Exploration, Nanjing 210093, China\\
}
\begin{document}
\date{Accepted. Received; in original form}
\pubyear{2018}

\maketitle

\label{firstpage}

\begin{abstract}
We have initiated a programme to study the physical/dynamical state of gas in galaxy clusters and the impact of the cluster environment on gaseous halos of individual galaxies using X-ray imaging and UV absorption line spectroscopy of background QSOs. Here we report results from the analysis \cha\ and \xmm\ archival data of five galaxy clusters with such QSOs, one of which has an archival UV spectrum. We characterize the gravitational masses and dynamical states, as well as the hot intracluster medium (ICM) properties of these clusters. Most clusters are dynamically disturbed clusters based on the X-ray morphology parameters, the X-ray temperature profiles, the large offset between X-ray peak and brightest cluster galaxy (BCG). The baryon contents in the hot ICM and stars of these clusters within $r_{500}$ are lower than the values expected from the gravitational masses, according to the standard cosmology. We also estimate column densities of the hot ICM along the sightlines toward the background QSOs as well as place upper limits on the warm-hot phase for the one sightline with existing UV observations. These column densities, compared with those of the warm and warm-hot ICM to be measured with UV absorption line spectroscopy, will enable us to probe the relationship among various gaseous phases and their connection to the heating/cooling and dynamical processes of the clusters. Furthermore, our analysis of the archival QSO spectrum probing one cluster underscores the need for high quality, targeted UV observations to robustly constrain the 10$^{5-6}$ K gas phase.
\end{abstract}


\section{Introduction}
\label{s:int}

Diffuse X-ray emission is routinely observed from the hot intracluster medium (ICM) of galaxy clusters. This emission arises chiefly 
in their inner regions ($r \lesssim r_{500}$, within which the mean mass 
density is 500 times the critical density of the Universe). 
Recently, however, diffuse X-ray emission has also been observed 
from outer regions of rich clusters as well (e.g., \citealt{2012NJPh...14b5010B}; 
\citealt{2012A&A...541A..57E}; \citealt{2013MNRAS.432..554W}; \citealt{2014MNRAS.439.1796W}). 
These observations indicate that the ICM is inhomogeneous, especially in outer regions, as predicted by theoretical 
models or simulations of structure formation (e.g., \citealt{2006MNRAS.373.1339R}; 
\citealt{2009ApJ...696.1640M}), which show a complicated shock 
heating/cooling history in the outer ICM regions (out to a few $r_{200}$, 
where the strong external shock of the accretion flow is located, e.g., 
\citealt{2009ApJ...696.1640M}) and an enhanced presence of the multiphase warm-hot plasma may be expected
(e.g., \citealt{2008MNRAS.385.1211P}). In addition, the cluster environment 
could strongly affect the circumgalactic medium (CGM) of individual 
galaxies via processes such as ram-pressure stripping and 
pressure compression (e.g. \citealt{2011MNRAS.413..347L}; \citealt{2015MNRAS.453.4051E}; \citealt{2015ApJ...806..103R}). 
However, the effectiveness of such stripping remains greatly uncertain, especially
in outer regions of clusters, as demonstrated in recent simulations, where the existence of
a complex pattern of flows, turbulence, and a continuous fueling of the CGM 
from the ICM is considered (e.g., \citealt{2017MNRAS.469...80Q}). New observational constraints are needed to make progress in our understanding of these phenomena and physical processes, which are important not only for determining the properties of the ICM 
and the effects on galaxy evolution, but also for properly using clusters as cosmology probes 
(e.g., via the Sunyaev-Zel'dovich effect).

We have been conducting a multi-wavelength study of how the properties of the ICM and CGM may be affected by the richness and dynamical state of galaxy clusters (e.g., \citealt{2014MNRAS.439.1796W}; \citealt{2016MNRAS.459..366G}; \citealt{2018MNRAS.475.2067B}). 
We use 1) X-ray observations to characterize the morphological and thermal properties of the hot ICM, 2) far-UV absorption-line spectroscopy to constrain the column densities, metal abundances, and kinematics of warm and warm-hot gas,
and 3) optical spectroscopy of galaxies to determine their 
associations with individual absorbers and X-ray-emitting substructures. 

Our initial studies were focused on a few optically selected clusters, 
as reported in \cite{2014MNRAS.439.1796W}, \cite{2016MNRAS.459..366G}, and \cite{2018MNRAS.475.2067B}. These studies are based on archival \cha\ data as well as 
new \xmm\ and {\em Hubble Space Telescope} (\hst)/Cosmic Origins Spectrograph (COS) observations. \cite{2014MNRAS.439.1796W} show how the density and temperature radial profiles (hence the hot gas properties of individual clusters along the QSO sightline) can be estimated even when the X-ray emission from the projected clusters is present. Based on \xmm\ observations,
\cite{2016MNRAS.459..366G} find that each of the two optically-selected clusters 
actually consists of distinct merging subcluster pairs at 
similar redshifts. These subclusters themselves typically 
show substantial substructures, including strongly distorted 
radio lobes, as well as large position offsets between the diffuse 
X-ray centroids and the brightest galaxies. Thus these clusters are dynamically
young systems. Comparing the hot gas and stellar masses of each cluster 
with the expected cosmological baryonic mass fraction indicates a 
significant deficit, which could be filled by other gas components 
\citep{2016MNRAS.459..366G}. 

\cite{2018MNRAS.475.2067B} present the {\sl HST}/COS study of
these clusters. They detect broad Ly$\alpha$ absorption (BLA) 
features associated with one cluster. However, rather than tracing material contained within the ICM, 
these features are consistent with metal-poor material in-falling 
from the intergalactic medium (IGM). Another QSO sightline
probes the interface region of a dynamically young system of 
merging subclusters and shows a quite narrow $b \sim 16 {\rm~km~s^{-1}}$
H I profile at the cluster system's redshift, which may represent a dense, 
cool ($T \sim 10^4$~K) cloudlet induced in the wake of the 
cluster merger shock. Interestingly, no OVI is detected to sensitive limits 
[$N$(OVI) $\lesssim 10^{13.7} {\rm~cm^{-2}}$]. While these results
are intriguing, it is still difficult to make any firm conclusions 
regarding the multi-phase ICM properties of galaxy clusters in general,
let alone their baryon contents. The sample of seven clusters
(including the newly discovered merging subclusters) that we 
have studied in these works is still too limited
to allow a meaningful statistical analysis. These clusters are dominated
by highly disturbed ones or even ongoing mergers. It is highly desirable 
to expand this sample to include more clusters that are
relatively isolated and relaxed and that, of course, have UV-bright background 
QSOs at various impact parameters so that
we can assess how the multi-phase ICM properties may depend on the cluster
dynamical state. 

Here, we present a study of a sample of relatively isolated 
{\sl X-ray-selected} clusters, which are paired with UV-bright 
background QSOs. This study focuses primarily on the analysis of 
archival \cha\ and \xmm\ data, representing a step 
to enable a statistically meaningful investigation of the multiphase 
ICM and CGM in and around clusters of galaxies.
One of the clusters' background QSOs has been observed with HST/COS, and we analyze this spectrum to constrain the baryon contribution of warm-hot 10$^{5-6}$ K gas to the cluster's baryon budget and demonstrate the need for targeted, high quality observations to address this scientific goal.
The rest of the present paper is organized as follows: In Section~\ref{s:obs} we describe
the sample selection and our data reduction procedures;
Section~\ref{s:res} presents the results based on the X-ray observations; 
Section~\ref{s:dis} discusses the dynamical state and baryon content of clusters.
Section~\ref{s:sum} summarizes our results.
We use the standard cold dark matter cosmology with $H_{0}=\rm{70\ km\ s^{-1}\ Mpc^{-1}}$, $\Omega_m$=0.3, and $\Omega_\Lambda$=0.7.

\section{Sample selection and X-ray data analysis}
\label{s:obs}

\subsection{Sample selection}
We find our sample pairs for the present study from cross-correlating
FUV-bright QSOs with a meta-catalog of X-ray-detected galaxy clusters \citep{2011A&A...534A.109P}. 
We select clusters with $0.1 \leq z_c \leq 0.4$ and $kT \geq 2$ keV, 
paired with UV-bright background QSOs of $m_{FUV} < 18.3$ and projected within $1.5 \times r_{200}$. In this redshift range, the warm and warm-hot ICM can be effectively observed with the COS to detect absorption lines of the O VI doublet, H I Ly$\alpha$ and Ly$\beta$, and a number of other ions. These lines uniquely constrain the thermal, kinematic, and chemical properties of cooler ICM/CGM gases at various cluster impact parameters. This redshift range is also ideal for weak lensing mapping of the gravitational mass distribution, Sunyaev-Zel'dovich effect measurements, multi-object galaxy spectroscopy, and X-ray imaging/spectroscopy of the hot ICM across the entire clusters. Here we focus on five galaxy clusters
for which good-quality archival \cha\ and \xmm\ observations are available 
(Table~\ref{t:sam}). None of these clusters have been carefully analyzed in the 
literature, except for inclusions in some large surveys focused on scaling relations and cosmology
(e.g., \citealt{2007A&A...469..363B}; \citealt{2013ApJ...767..116M}; \citealt{2015MNRAS.449..199M}).


\begin{table}
 \centering
  \caption{\cha\ and \xmm\ observations}
  \begin{tabular}{@{}lccc@{}}
\hline\hline
Name & Obs-ID & Exp (ks) & Clean Exp (ks)\\ 
\hline
J0350 & 7227 (ACIS-I) & 24.7 & 18.6\\   
A655 & 15159 (ACIS-I) & 8.0 & 7.2\\
A959 & 0406630201 (EPIC)  & 25.6/23.8/23.8$^a$ & 3.3/7.9/8.1\\
A1084 & 0201901501 (EPIC) & 25.2/29.2/29.2 & 17.4/23.9/24.1\\
A2813 & 0042340201 (EPIC) & 10.0/14.4/14.4 & 5.7/10.7/10.7\\  
\hline
\end{tabular}
\begin{tablenotes}
\item
 $^a$EPIC exposures of the pn/MOS1/MOS2 cameras.
\end{tablenotes}
\label{t:sam}
\end{table}

\subsection{\cha\ data}
Two clusters were observed with the \cha/ACIS-I (Table~\ref{t:sam}).
We use \cha\ Interactive Analysis of Observation (CIAO, version 4.7) and calibration database (CALDB, version 4.6.9) to reprocess the \cha\ data, following a procedure
similar to that detailed in \cite{2014MNRAS.439.1796W}. For each observation, we first reprocess a new level = 2 event file, using {\tt chandra\_repro} script with VFAINT mode correction, and then clean  time intervals strongly affected by flares, using the {\tt deflare} script. 
Such intervals are defined to be first deviating more than 3$\sigma$ and then a factor of $\geq$ 1.2 from the mean rate. The light curve is extracted from a bright source-free region in the 2.3-7.3 keV band, which is most sensitive to flares due to both the spectral shapes of the flaring and the minimum quiescent instrument plus sky  backgrounds \citep{2006ApJ...645...95H}. 
The exposure times of the clean data, as well as the original ones, 
are included in Table~\ref{t:sam}.
We use {\tt wavdetect} to detect discrete sources. The standard stowed background events
are reprojected to match each of the two ACIS-I observations. The event rate 
of the background is rescaled to the count rate in the 9-12 keV band, where the \cha\ effective area is negligible and the flux is dominated by the particle background. 

We construct the count, stowed background, and effective exposure map in the 0.5-2.0 keV band, where the X-ray emission is dominated by the clusters. The background subtracted and 
exposure corrected intensity maps are then generated and smoothed 
with a 10$^{\prime\prime}$ sigma Gaussian kernel using {\tt aconvolve}. 

The spectra are extracted with {\tt specextract}. For the spectral fitting, we firstly subtract the instrumental background, and secondly subtract the sky background. The instrumental non-X-ray background spectra for the on-cluster spectra are estimated with the rescaled stowed data in the same detector regions. For the diffuse X-ray sky contribution, we use the HEASARC X-ray Background Tool to extract an off-cluster {\em ROSAT} all-sky survey (RASS) spectrum in an annulus (of inner and outer radii equal to 1-2 degree) around the cluster. We then jointly fit the non-X-ray background-subtracted on- and RASS off-cluster spectra. The off-cluster spectrum is fitted with a
3-component model consisting of an unabsorbed thermal model {\tt apec} (for the Local Bubble
contribution), an absorbed {\tt apec} (the Galactic halo gas)  and an 
absorbed {\tt power-law} (the extragalactic background). In addition to 
this 3-component model, an additional absorbed {\tt apec} 
(representing the cluster emission) is included to fit the on-cluster spectrum.
Fig.~\ref{fig:bkg} shows the relation between derived on-cluster soft sky flux and off-cluster RASS flux, they are correlated as expected \citep{2009ApJ...693.1142S}.

\subsection{\xmm\ data}
The \xmm\ data used here are from the European Photon Imaging 
Camera (EPIC) consisting of two MOS and one pn CCD arrays.
We process the  data using the Extended Source Analysis Software (ESAS; \citealt{2008A&A...478..575K}; \citealt{2008A&A...478..615S}), integrated into the \xmm\ Science Analysis System (SAS, version 13.5.0.) with the associated Current Calibration Files (CCF), following the procedures detailed in \cite{2016MNRAS.459..366G}.
Briefly, we use {\tt emchain} and {\tt epchain} to reproduce the event files from MOS and pn CCDs. The flares are filtered out with {\tt mos-filter} and {\tt pn-filter}. The point sources are detected by {\tt cheese}. 

We apply the {\tt mos-spectra} and {\tt pn-spectra} to create spectra and images. The instrumental background is modeled with routines {\tt mos\_back} and {\tt pn\_back}, which use data from
the unexposed pixels in the detector corner, and filter-wheel closed data sets with hardness ratios and count rates similar to those measured during the observations.

The MOS1, MOS2, and pn images are combined with the {\tt comb}, then we use the routine {\tt adapt} to create the quiescent instrumental background subtracted, exposure corrected EPIC images, which are binned by a factor of 2 and adaptively smoothed with a minimum number of 100 counts bin$^{-1}$.

Similarly, we use the HEASARC X-ray Background Tool to extract an off-cluster RASS sky background spectrum, which is also fitted with the 3-component model. 
Then the quiescent instrumental background-subtracted MOS1, MOS2 and pn on- and RASS off-cluster spectra are jointly fitted. However, here we need to include additional 
model components for the residual instrumental and/or variable sky
background: 1) {\tt gaussian} emission lines to account for a few strong instrumental 
lines and for possible solar wind charge exchange contributions 
to the OVII and OVIII K$\alpha$ lines and 2) a {\tt power-law} 
not folded through the instrumental effective areas for residual soft proton contamination \citep{2008A&A...478..575K}.

\section{Results} 
\label{s:res}

We here describe the general X-ray morphological and spectral properties
of the clusters, while their individual multiwavelength 
characteristics will be detailed in \S~\ref{ss:multi}.

\subsection{Morphological Structure}

\begin{figure*}
\includegraphics[width=0.455\textwidth,keepaspectratio=true,clip=true]{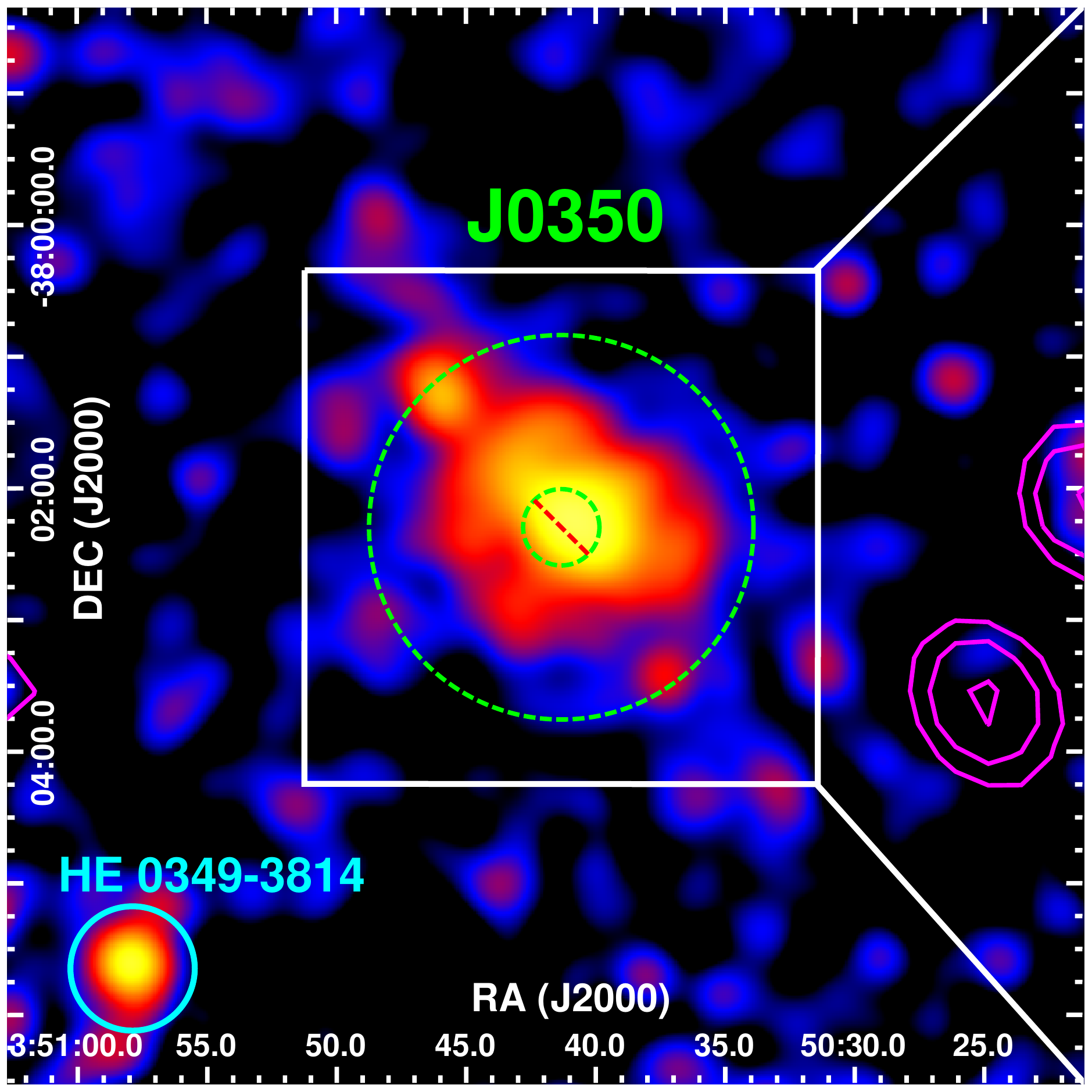}
\includegraphics[width=0.455\textwidth,keepaspectratio=true,clip=true]{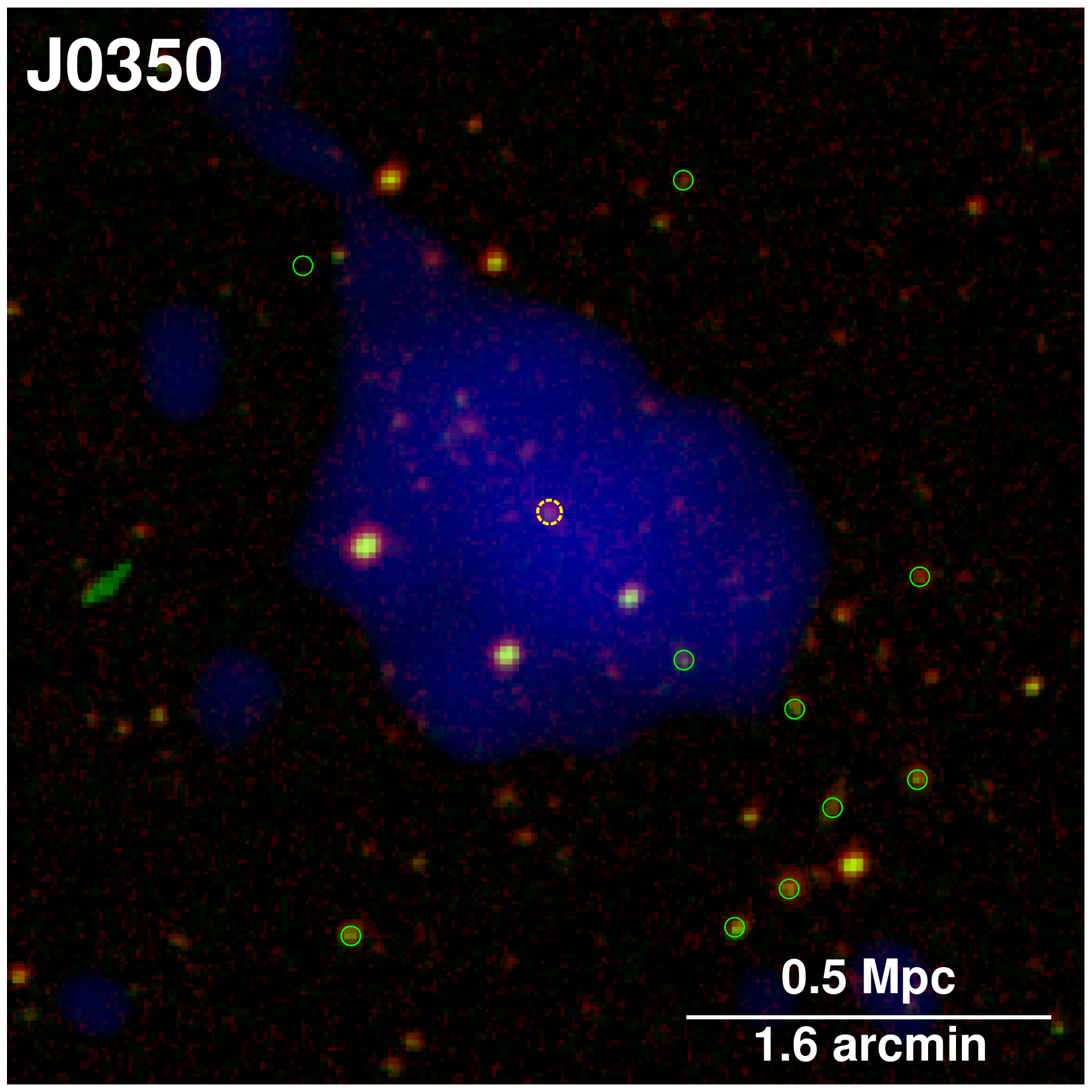}
\includegraphics[width=0.455\textwidth,keepaspectratio=true,clip=true]{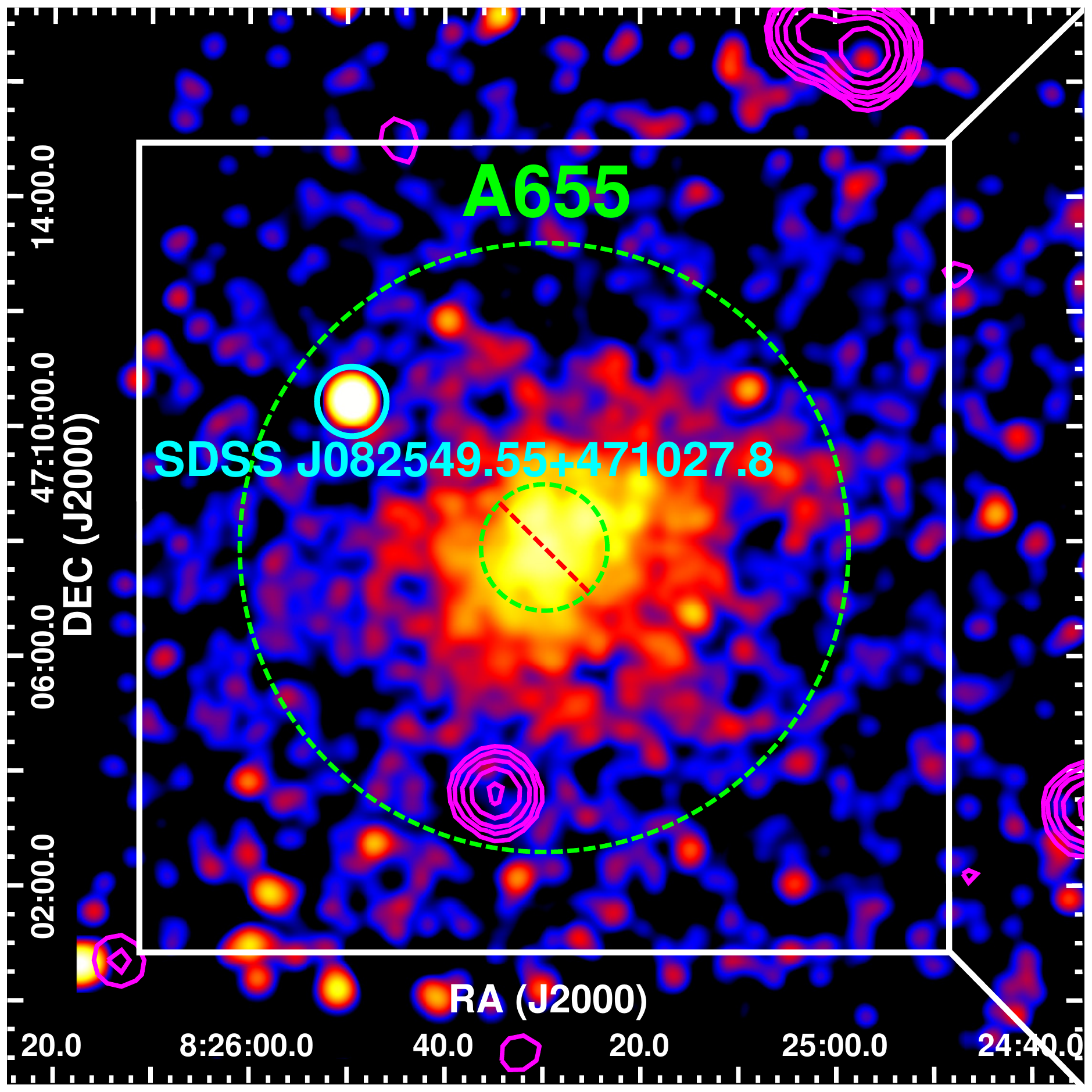}
\includegraphics[width=0.455\textwidth,keepaspectratio=true,clip=true]{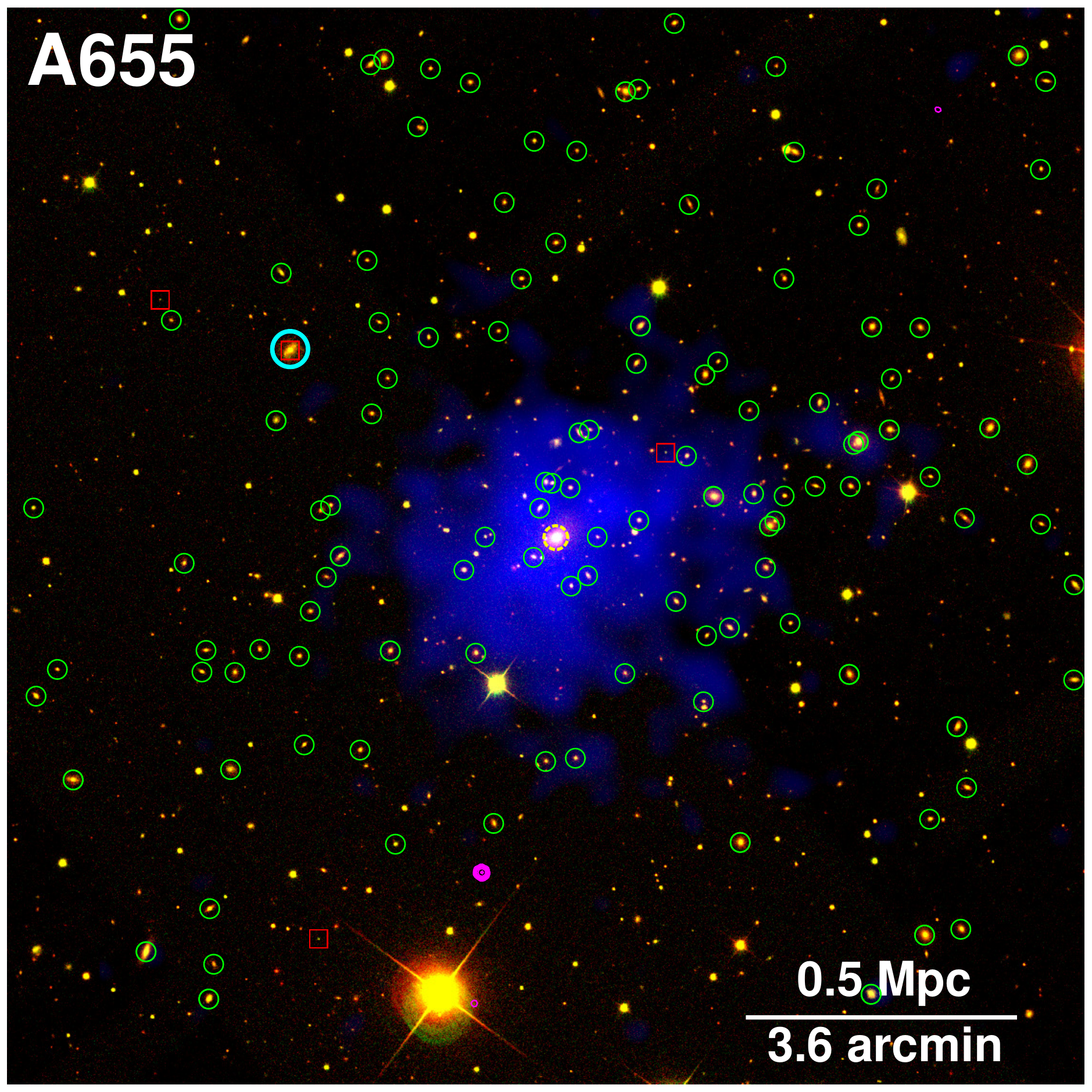}
\includegraphics[width=0.455\textwidth,keepaspectratio=true,clip=true]{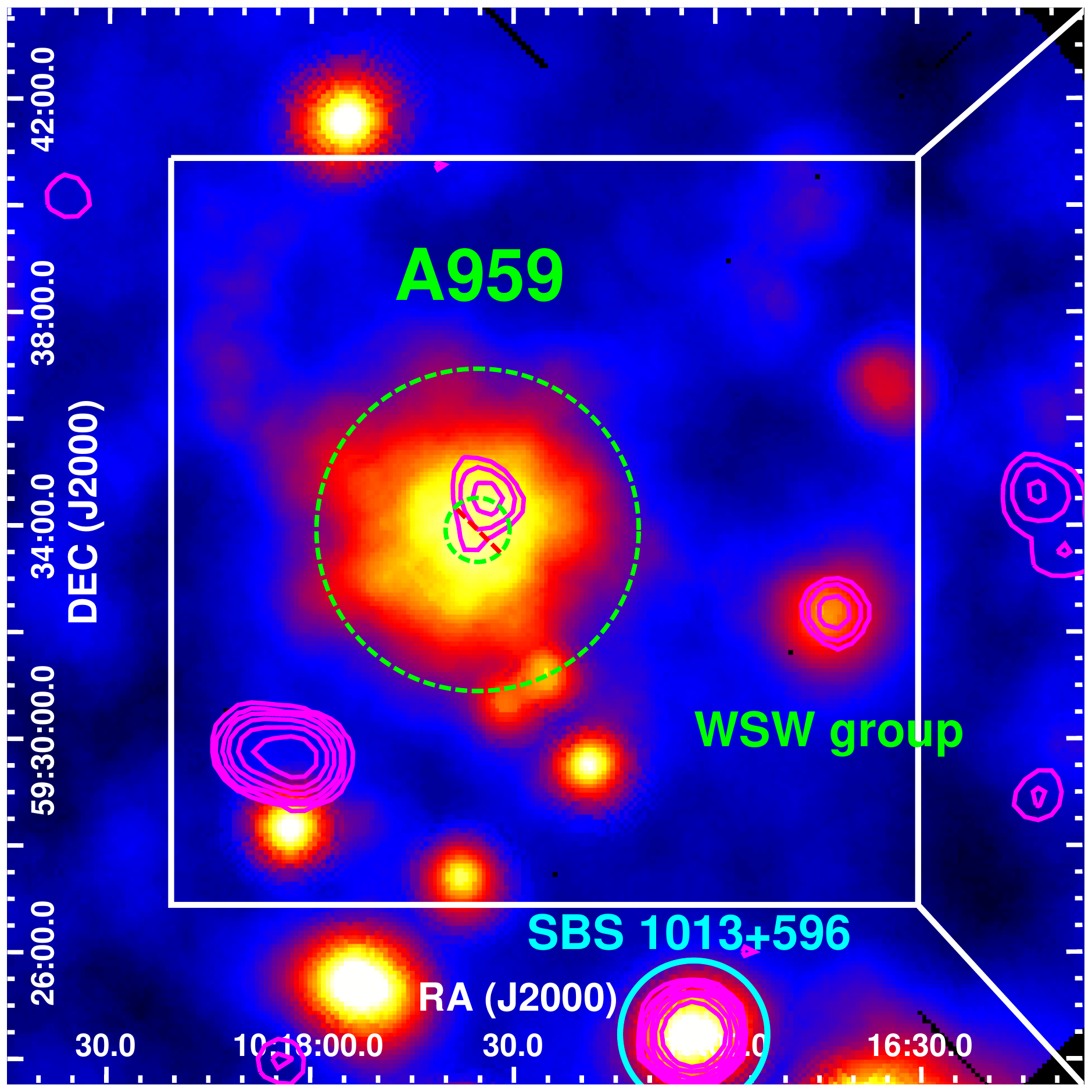}
\includegraphics[width=0.455\textwidth,keepaspectratio=true,clip=true]{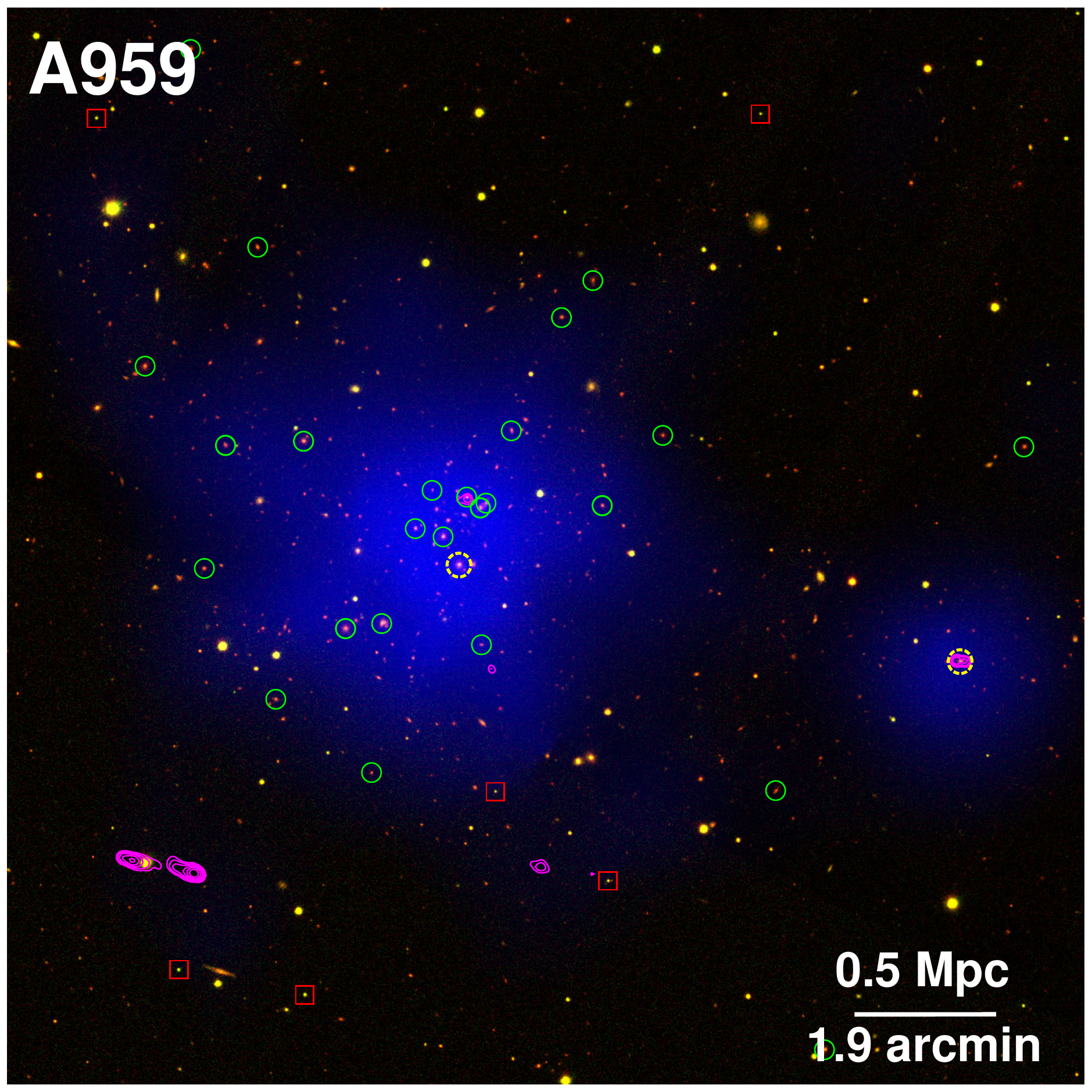}
\end{figure*}

\begin{figure*}
\includegraphics[width=0.455\textwidth,keepaspectratio=true,clip=true]{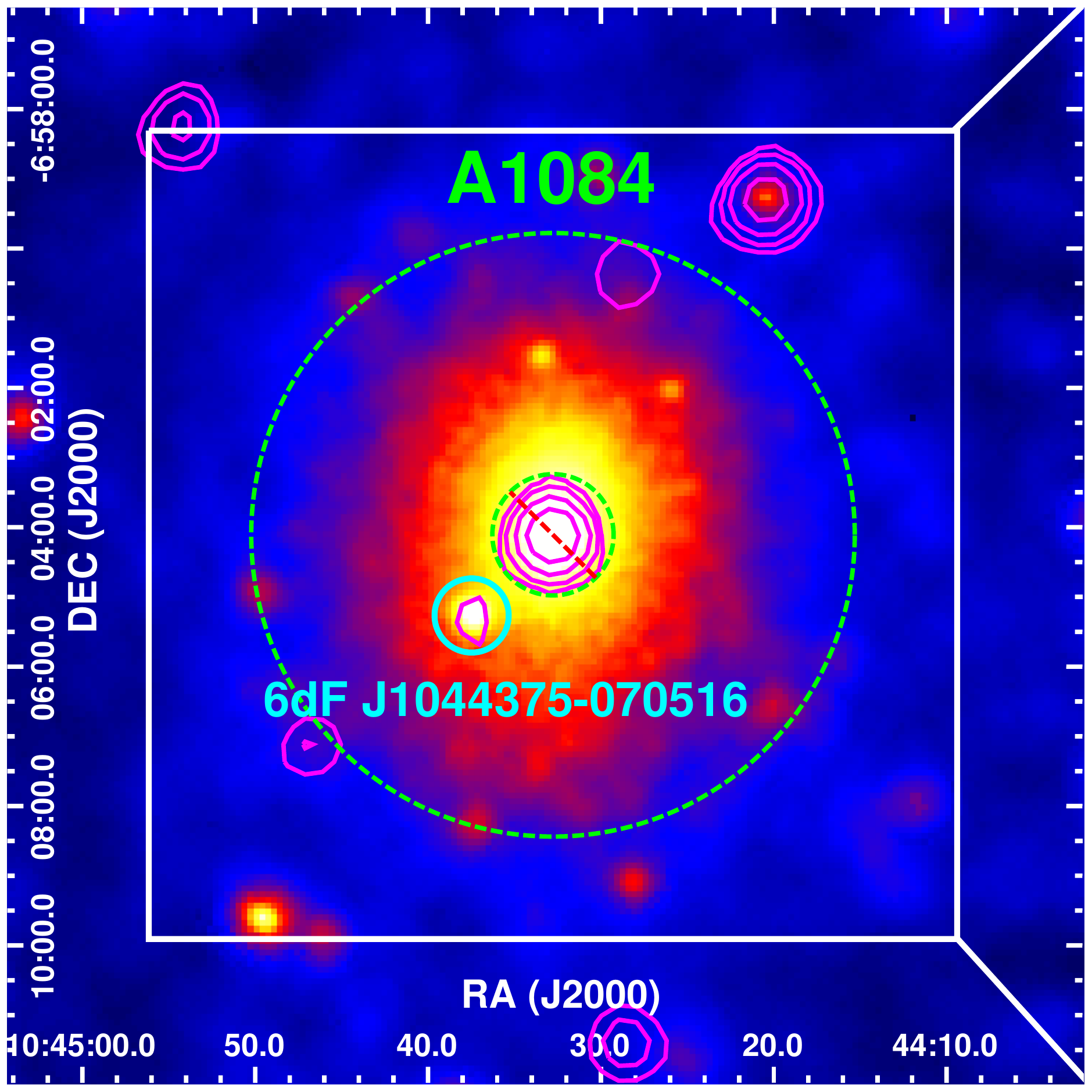}
\includegraphics[width=0.455\textwidth,keepaspectratio=true,clip=true]{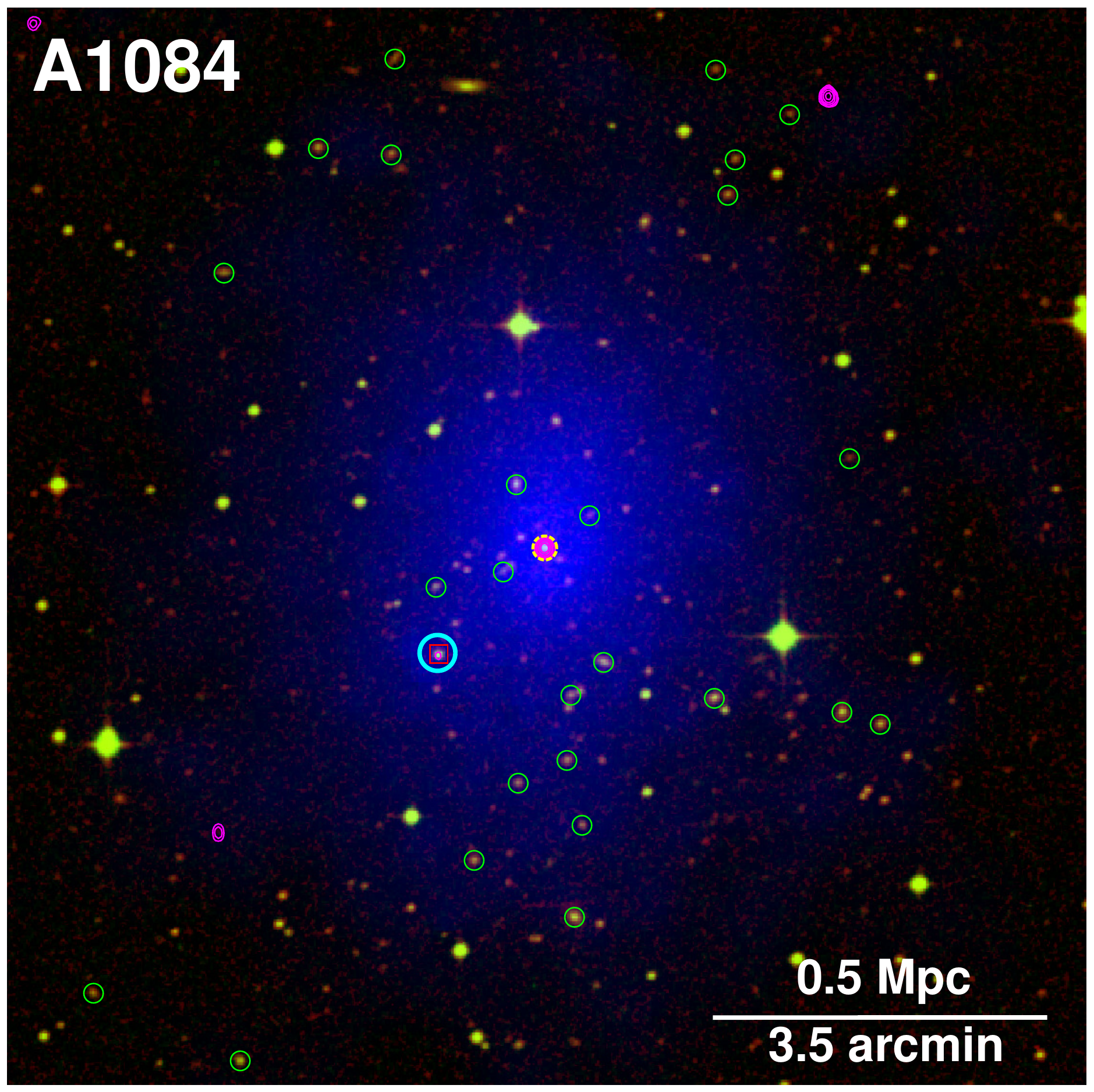}
\includegraphics[width=0.455\textwidth,keepaspectratio=true,clip=true]{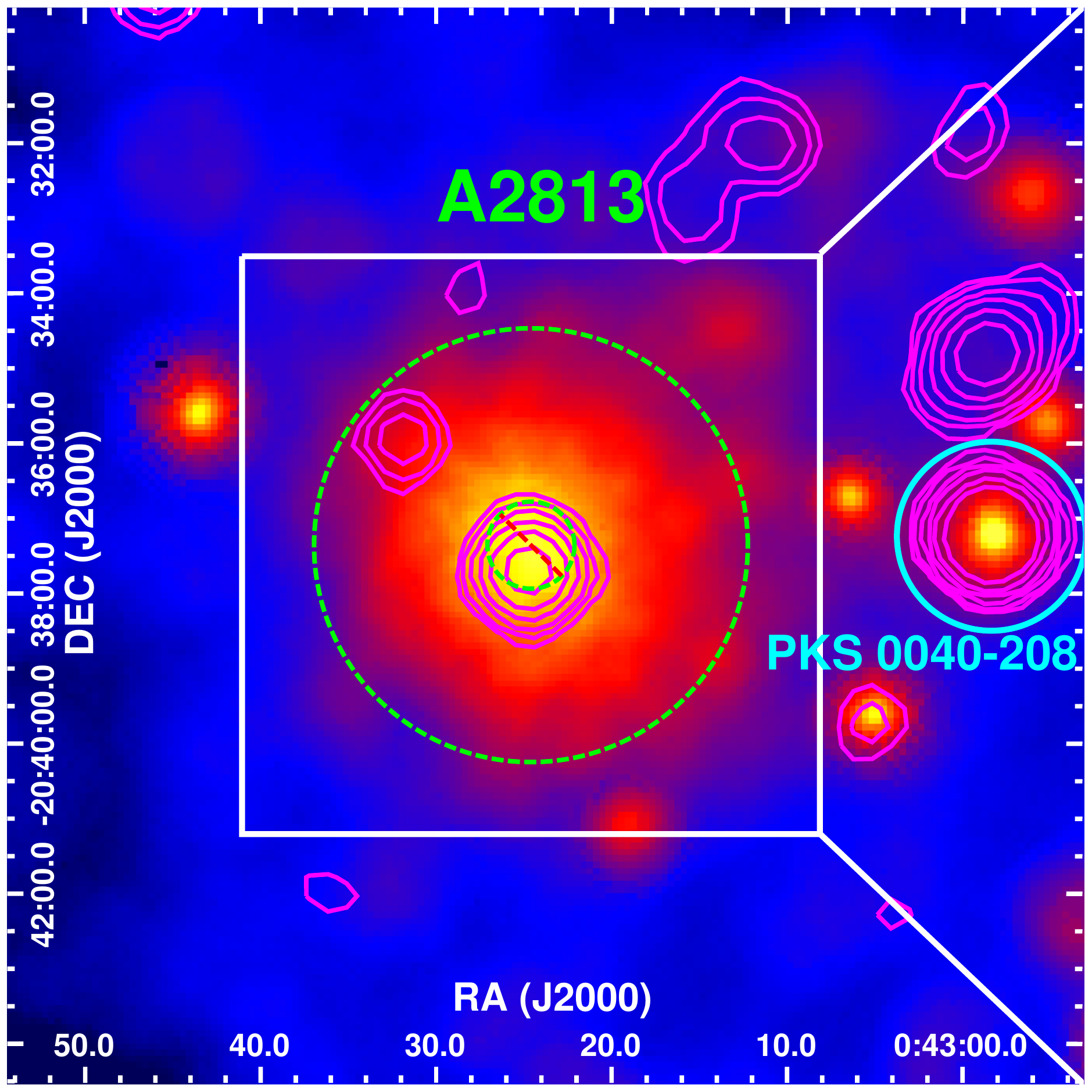}
\includegraphics[width=0.455\textwidth,keepaspectratio=true,clip=true]{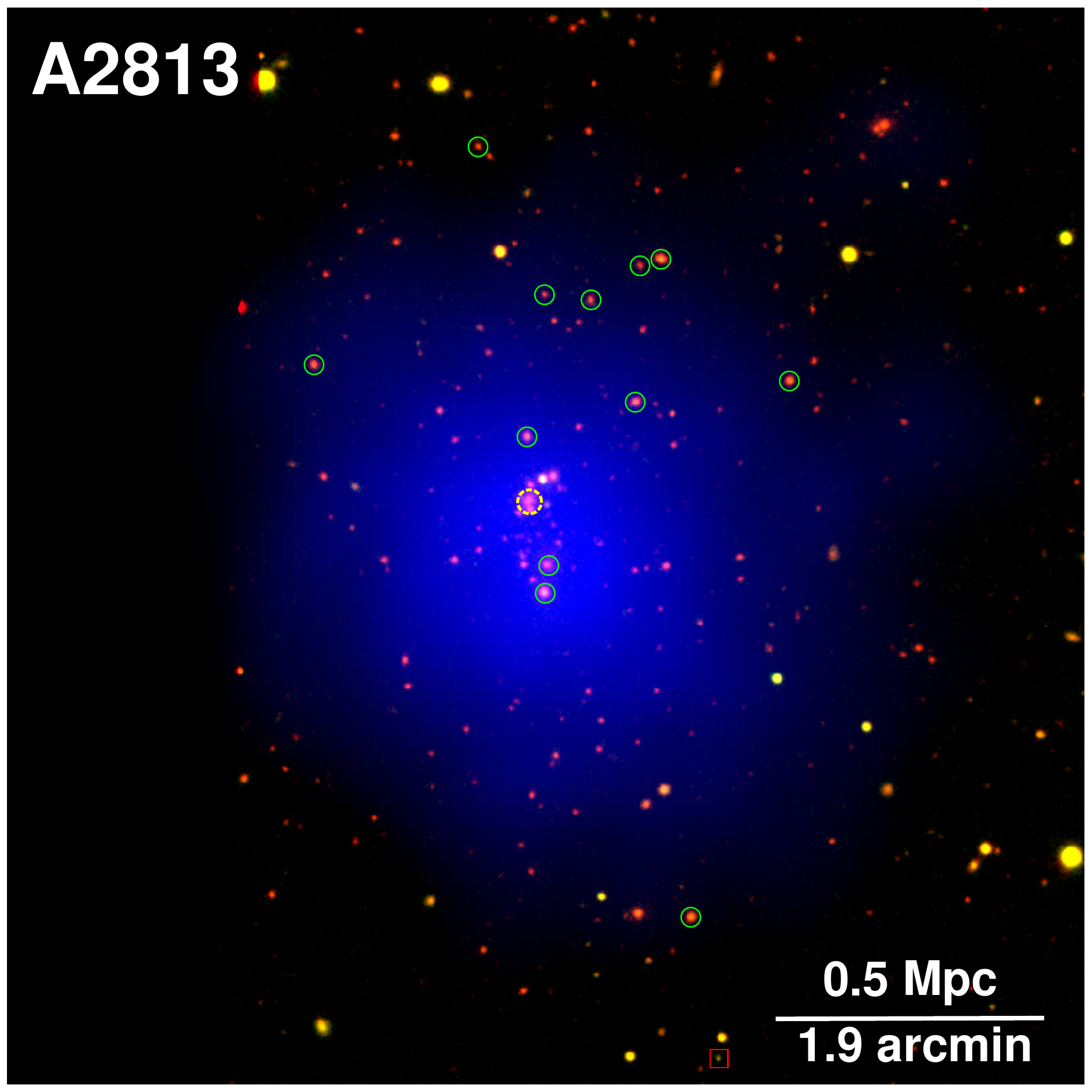}
	\caption{
{\sl Left}: \cha/\xmm\ 0.5-2 keV background subtracted, exposure corrected, and smoothed image, overlaid with 1.4 GHz continuum magenta contours from NVSS at 2, 4, 8, 16, 32, and 64 mJy beam$^{-1}$. The UV-bright QSO is marked as a cyan circle and labeled. The white box outlines a close-up region, and the dashed annulus is the 0.15-0.75 $r_{500}$ region for global temperature evaluation.
{\sl Right}: a close-up multiwavelength montage: SDSS r-band (red), SDSS g-band (green), and diffuse 0.5-2 keV emission with point sources excluded (blue); DSS red band (red), DSS blue band (green) for J0350 and A1084. Galaxies (solid green circles), BCGs (dashed yellow circles), and QSOs (red boxes, UV-bright ones are marked with additional cyan circle) are labeled. The galaxy location are from SDSS (galaxy redshift within 0.01 of BCG from spectroscope database and 0.03 from photometric database for A655, A959, and A2813) or NED (J0350 and A1084). Overlaid magenta contours represent the 1.4 GHz continuum intensity at 1, 2, 4, 8, 16, and 32 mJy beam$^{-1}$
from the FIRST survey (except for J0350 and A2813).}
	 \label{fig:img}
\end{figure*}

Fig.~\ref{fig:img} presents the X-ray images 
of the cluster/QSO pair fields. The large fields of view shown in the
left panels include the UV-bright background QSOs, which are also luminous in
 X-ray and are marked with cyan circles. The right panels show
the multiwavelength close-ups of the clusters, including the
diffuse X-ray emission intensity maps 
with detected discrete sources excised, as well as optical images and
radio continuum contours.
Each field contains one known cluster, except for A959, which is 
accompanied by a small group at its west-southwest (WSW) direction.
A more quantitative discussion of these X-ray morphology will be given in \S~\ref{ss:dyn}.

\begin{figure*}
\includegraphics[width=0.195\textwidth,keepaspectratio=true,clip=true]{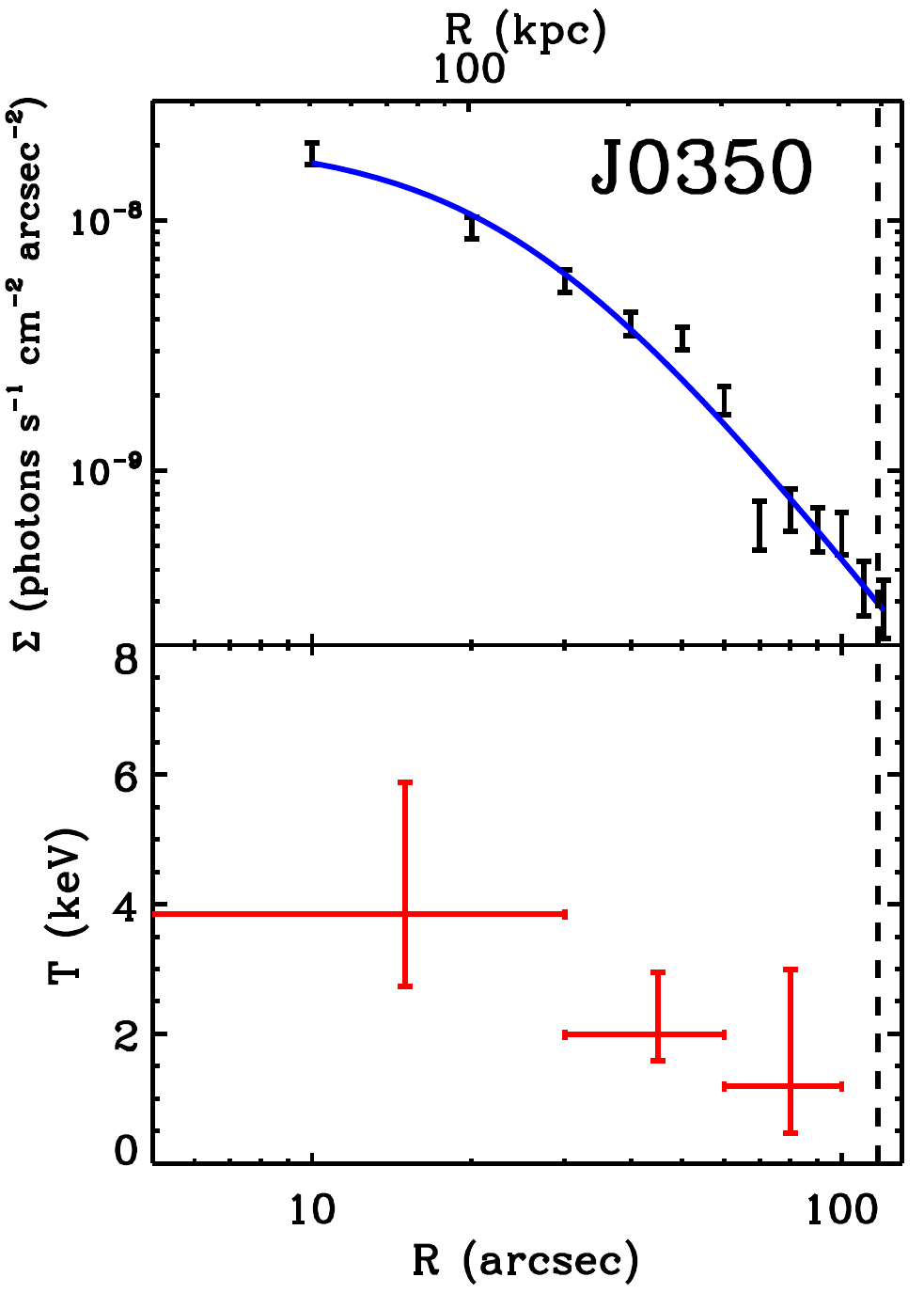}
\includegraphics[width=0.195\textwidth,keepaspectratio=true,clip=true]{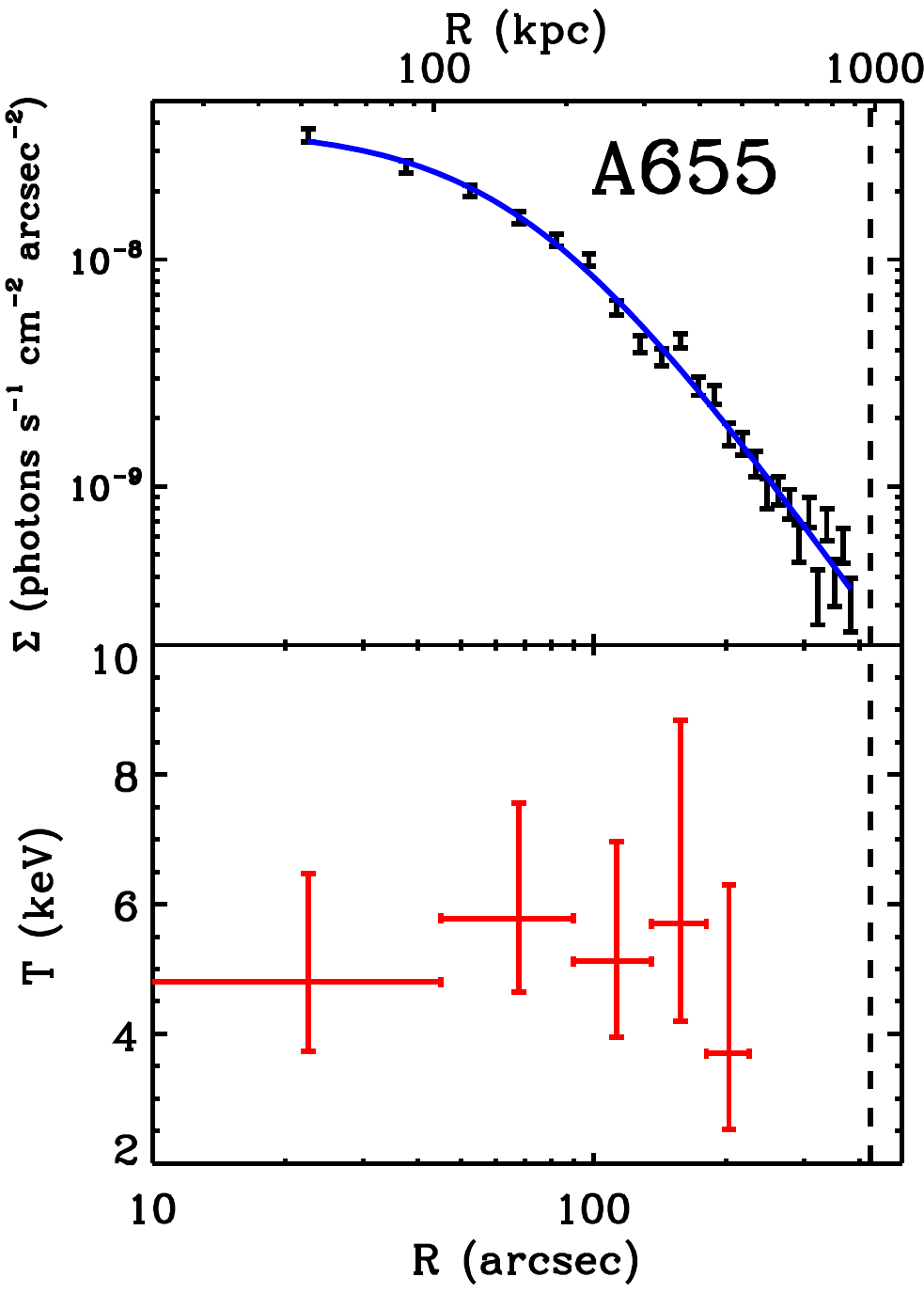}
\includegraphics[width=0.195\textwidth,keepaspectratio=true,clip=true]{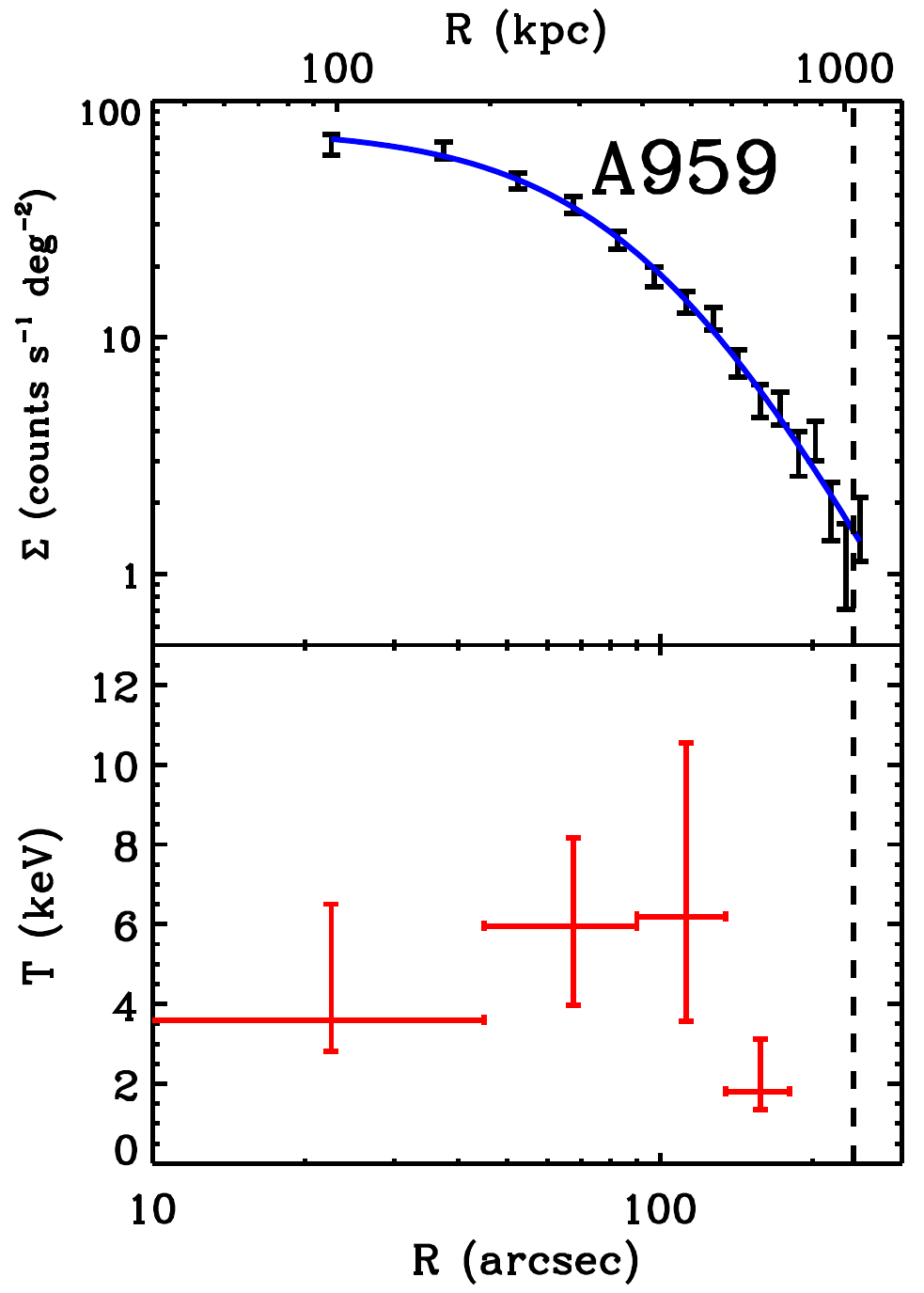}
\includegraphics[width=0.195\textwidth,keepaspectratio=true,clip=true]{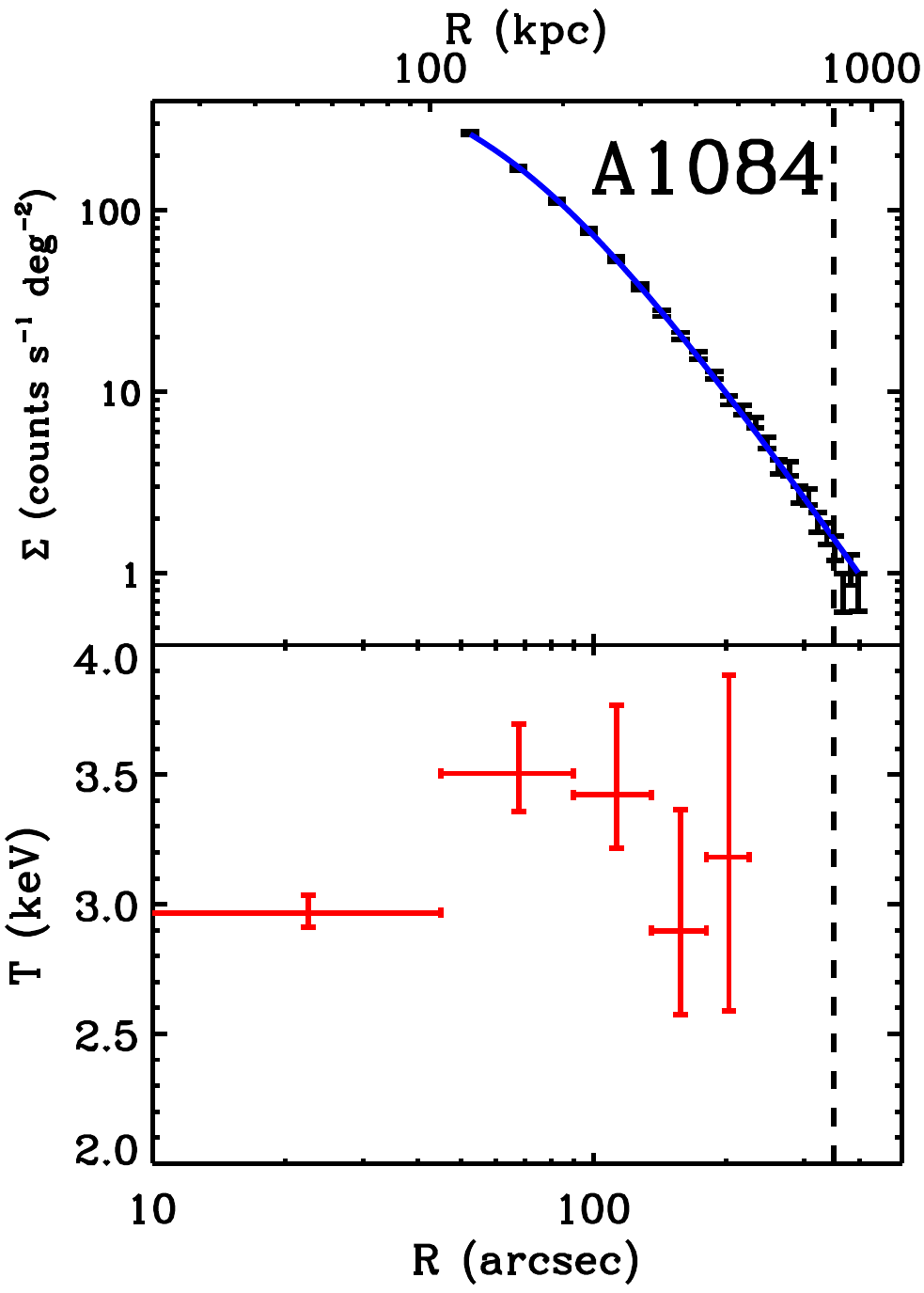}
\includegraphics[width=0.195\textwidth,keepaspectratio=true,clip=true]{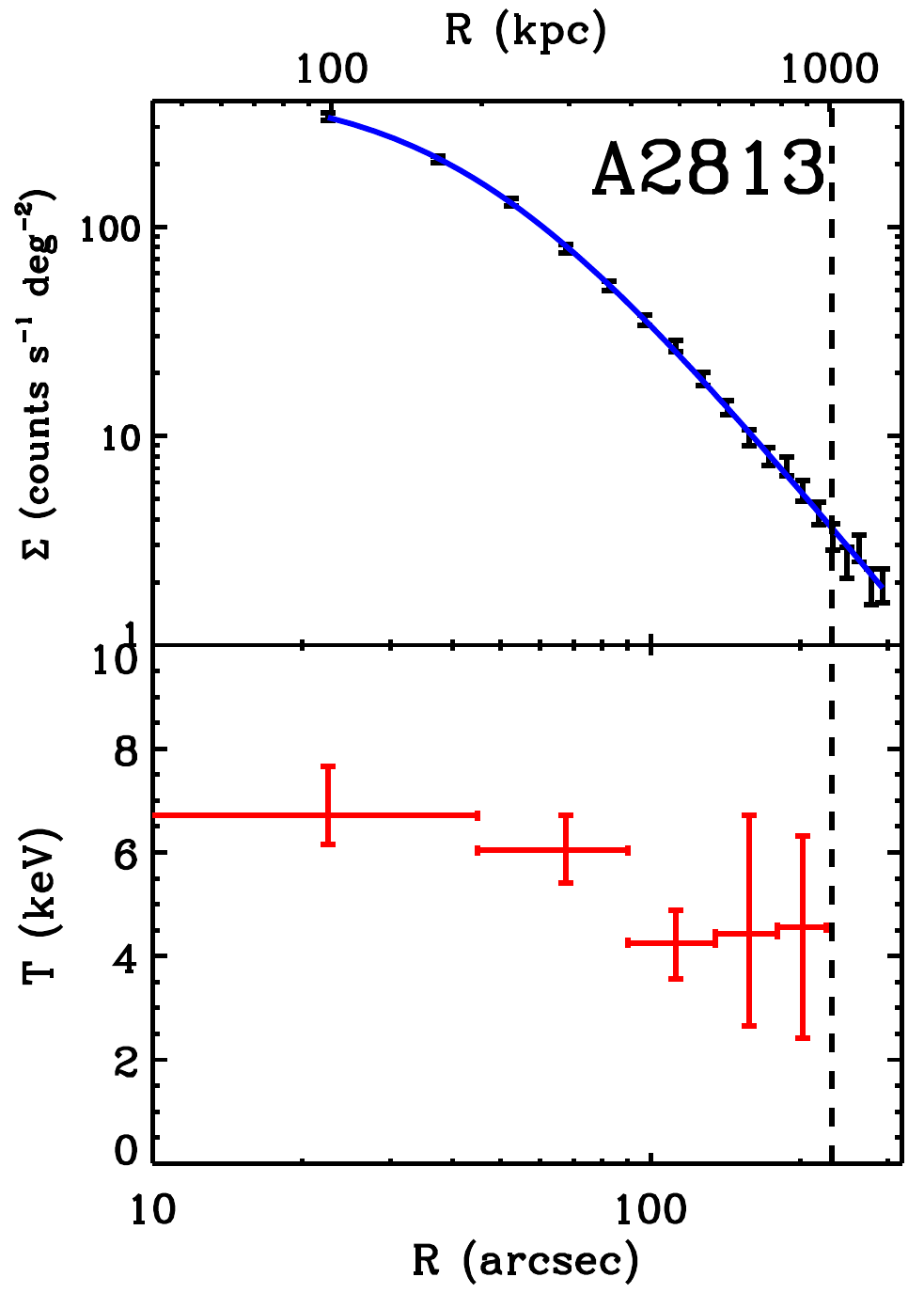}
	\caption{{\it Top panels}:  Radial intensity profiles of the diffuse 0.5-2 keV emission for our sample clusters, together with the respective best-fitting $\beta$-models. The dashed line shows the location of $r_{500}$. {\it Bottom panels}: Radial temperature profiles of the clusters.}
	 \label{fig:RP}
\end{figure*}

The upper panels of Fig.~\ref{fig:RP} presents the radial intensity profile of the diffuse 
X-ray emission. 
These profiles are constructed around the X-ray peaks of individual 
clusters and are fitted with the 
standard $\beta$-model \citep{1976A&A....49..137C} of the form: 
\begin{equation}
I=I_{0} (1+x^{2})^{1/2-3\beta},
\end{equation}
where $x=R/r_{c}$, while $I_{0}$, $r_c$, and $\beta$ are the free parameters.
Their best-fitting values are included in Table~\ref{t:clusters}.
The profiles reach out to $r_{500}$, except for A655 due to its relatively shallow data.
The $\beta$ value is consistent with the results of \cite{2003MNRAS.340..989S}, who find $\beta$ is close to canonical value of 2/3 for hot ($>3-4$ keV) clusters and becomes increasing flattened below this temperature.
We use the fit result to estimate the central gas density and $M_{gas}$ within $r_{500}$ in \S~\ref{ss:mgas}.

\subsection{Thermal Properties}
The cluster emission is fitted with the {\tt XSPEC} (version 12.8.0; \citealt{1996ASPC..101...17A}) package and the AtomDB (version 2.0.1) database of atomic data. We use {\tt apec} emissivity model to fit the on-cluster emission, with the model metallicity fixed to 0.3 solar \citep{2009ARA&A..47..481A} and redshift fixed to the optical redshift of BCG. We also apply the Tuebingen-Boulder absorption model ({\tt tbabs}) for
X-ray absorption by the interstellar medium (ISM), with the hydrogen column density $N_{\rm H}$ fixed to the Galactic value from the NHtot tool \citep{2013MNRAS.431..394W}.
The spectroscopic X-ray temperature is measured in 0.15-0.75 $r_{500}$ (marked as dashed annuli in the left panels of Fig.~\ref{fig:img}), the inner boundary of 0.15$r_{500}$ is chosen to exclude most of the possible cool core (CC) region with a large dispersion in observation, while the outer boundary of 0.75 $r_{500}$ is limited by the quality of the spectroscopic data. The $r_{500}$ is estimated iteratively from $M-T$ relation (\citealt{2009ApJ...693.1142S} for \cha\ calibration and \citealt{2007A&A...474L..37A} for \xmm\ calibration). Table~\ref{t:clusters} includes the results of our spectral characterization of the
hot ICM in the individual clusters. The single-temperature plasma model
gives acceptable fits to all of them.

We also produce the radial temperature profiles of these clusters in the bottom panels of Fig.~\ref{fig:RP}. 
Only A1084 shows a strong temperature drop toward the center when considering the error bars in temperature, which indicates that A1084 is a strong CC cluster.

\begin{table*}
\caption{Properties of galaxy clusters }
\tabcolsep=0.10cm
\footnotesize
\begin{tabular}{lccccc}
\hline\hline
Name & J0350 & A655 & A959 & A1084 & A2813\\
\hline
Redshift  & 0.363 & 0.129 & 0.288 & 0.133 & 0.292\\
X-ray peak position		&	$3^h50^m40.8^s, -38^\circ02^{\prime}14^{\prime\prime}$	&	$8^h25^m29.8^s, +47^\circ07^{\prime}57^{\prime\prime}$	&	$10^h17^m35.8^s, +59^\circ34^{\prime}03^{\prime\prime}$	&	$10^h44^m33.1^s, -7^\circ04^{\prime}11^{\prime\prime}$	&	$0^h43^m24.6^s, -20^\circ37^{\prime}27^{\prime\prime}$\\
BCG position &	$3^h50^m41.3^s, -38^\circ02^{\prime}09^{\prime\prime}$		&	$8^h25^m29.1^s, +47^\circ08^{\prime}01^{\prime\prime}$	&	$10^h17^m34.3^s, +59^\circ33^{\prime}40^{\prime\prime}$	&	$10^h44^m32.9^s, -7^\circ04^{\prime}08^{\prime\prime}$	&	$0^h43^m25.1^s, -20^\circ37^{\prime}02^{\prime\prime}$	\\
Offset (kpc)	&	38.4&	18.7	&	112.6	&	12.6	&	88.7		\\
Temperature (keV)&	$2.16^{+0.82}_{-0.41}$		&	$4.25^{+0.84}_{-0.54}$	&	$5.87^{+0.80}_{-1.33}$	&	$3.31^{+0.12}_{-0.13}$	&	$5.55^{+0.99}_{-0.44}$	\\
$f_x$ or $CR$&	0.79	&	9.68 	&	0.12 	&	0.41 	&	0.24	\\
$\eta (10^{-4})$ &  3.5 & 39.1 & 16.7 & 49.8 & 34.9\\
$\chi_{T}^{2}/d.o.f.$ &	22/23		&	114/111	&	306/300	&	1603/1631 &	766/849\\
$M_{500} (10^{14} M_\odot)$	&	$0.87^{+0.55}_{-0.28}$		& $3.00^{+1.00}_{-0.65}$	&	$4.36^{+1.03}_{-1.70}$	&	$1.78^{+0.14}_{-0.15}$	&	$3.96^{+1.22}_{-0.56}$\\
$r_{500}$ (Mpc)&	$0.59^{+0.12}_{-0.06}$		&	$0.98^{+0.11}_{-0.07}$	&	$1.04^{+0.08}_{-0.14}$	&	$0.82^{+0.02}_{-0.02}$	&	$1.01^{+0.10}_{-0.05}$	\\
$I_0$&	$2.1^{+0.6}_{-0.4}$	&	$3.8^{+0.4}_{-0.3}$	&	$76.3^{+9.7}_{-8.6}$	&	$670.4^{+67.5}_{-57.3}$	&	$465.7^{+43.4}_{-38.8}$		\\
$r_c$ (Mpc)&	$0.12^{+0.04}_{-0.03}$		&	$0.16^{+0.02}_{-0.02}$	&	$0.43^{+0.11}_{-0.08}$	&	$0.15^{+0.01}_{-0.01}$	&	$0.19^{+0.02}_{-0.02}$	\\
$\beta$&	$0.62^{+0.09}_{-0.06}$		&	$0.62^{+0.03}_{-0.03}$	&	$0.85^{+0.18}_{-0.12}$	&	$0.75^{+0.02}_{-0.02}$	&	$0.65^{+0.02}_{-0.02}$	\\
$\chi_{\beta}^{2}/d.o.f.$&	27.3/9		&	47.6/22	&	7.5/13	&	19.0/21	&	6.7/16\\
$n_0$ (10$^{-3}$ cm$^{-3}$) 	&	$2.54^{+0.10}_{-0.11}$	&	$2.87^{+0.11}_{-0.12}$	&	$1.46^{+0.05}_{-0.05}$ 	&	$6.60^{+0.23}_{-0.26}$ 	&	$4.92^{+0.18}_{-0.20}$	\\
$M_{gas} (10^{13} M_\odot)$&	$0.80^{+0.14}_{-0.12}$	&	$2.73^{+0.58}_{-0.48}$	&	$4.32^{+0.51}_{-0.46}$	&	$2.68^{+0.65}_{-0.52}$ &	$5.91^{+1.17}_{-0.98}$ 		\\
$M_{star} (10^{12} M_\odot)$		&	2.98&	5.67	&	6.88	&	4.32	&	6.55	\\
$D_{\rm QSO}$ (Mpc)&	1.45	&	0.58 	&	2.68 	&	0.23	&	1.61 		\\
$N_p$ ($10^{20}$ cm$^{-2}$) &	$4.3^{+3.5}_{-1.8}$	&	$15.9^{+7.3}_{-4.6}$ 	&	$2.7^{+2.0}_{-1.1}$ 	&	$36.9^{+9.9}_{-7.2}$  	&	$12.6^{+9.0}_{-5.0}$\\
\hline 
\end{tabular}
\begin{tablenotes}
      \small
      \item $Note.$ Errors are at the 90\% confidence level. The offset is the projected distance between
the X-ray peak and the BCG. The photon flux $f_x$ $(\rm{10^{-4}\ photons\ s^{-1}\ cm^{-2}})$ for \cha\ data or count rate CR $(\rm{count\ s^{-1}})$ for \xmm\ data, and $\eta$ is {\tt apec} normalization, as well as the temperature, are from the best-fitting {\tt apec} model of each cluster. The cluster mass $M_{500}$ and radius $r_{500}$ are estimated from the $M-T$ relation (\citealt{2009ApJ...693.1142S} for \cha\ calibration and \citealt{2007A&A...474L..37A} for \xmm\ calibration). The $\beta$-model parameters of Eq. (1) are from the best fit  to the 0.5-2 keV intensity profile, the $I_0$ is in the unit of $\rm{10^{-8}\ photons\ s^{-1}\ cm^{-2}\ arcsec^{-2}}$ for \cha\ data or $\rm {count\ s^{-1}\ deg^{-2}}$ for \xmm\ data. The inferred parameters include the central proton (hydrogen) density ($n_0$), the hot gas mass ($M_{gas}$) and stellar mass ($M_{star}$) within $r_{500}$, the stellar mass is estimated with the relation of \citep{2013ApJ...778...14G}. The proton column density ($N_p$) is at the projected distance ($D_{\rm QSO}$) of the corresponding UV-bright background QSO.
    \end{tablenotes} 
\label{t:clusters}    
\end{table*}

\section{Discussion}
\label{s:dis}
\subsection{Comparison with existing multiwavelength observations}
\label{ss:multi}

We here present a brief comparison of the above X-ray results with 
the existing multiwavelength observations, focusing on the properties relevant to the dynamical states (more detailed in \S~\ref{ss:dyn}) of the individual clusters. The right panels of Fig.~\ref{fig:img} show the optical image: SDSS for A655, A959, and A2813; DSS for  J0350 and A1084.  We also show the radio intensity contours from the NRAO VLA Sky Survey (NVSS; \citealt{1998AJ....115.1693C}) in the left panels of Fig.~\ref{fig:img} and those from the Faint Images of the Radio Sky at Twenty-cm (FIRST; \citealt{1995ApJ...450..559B}) in the right panels, except for  J0350 and A2813, which are not covered by FIRST. We next discuss some notable features for the individual cluster.

{\sl J0350.} Data in other bands are not notable for this cluster, although we notice a filament-like structure in the NE of cluster from the X-ray image.

{\sl A655.} It is classified as a BM I-type \citep{1970ApJ...162L.149B} cluster for its dominant cD galaxy.
It does not host any significant diffuse radio emission \citep{2009ApJ...697.1341R}.
The large core radius ($\sim$1300 kpc; \citealt{2005MNRAS.359..191S}) from a $\beta$-model fitting of member galaxy distribution may indicate it as a disturbed cluster, because merging events could have erased the original cusp profile. 

{\sl A959.} This cluster has been studied extensively. Most notable are the mapping of
gravitational mass via weak gravitational lensing observations
\citep{2003ApJ...591..662D} and the optical spectroscopy for individual galaxies 
\citep{2009A&A...495...15B}. In these studies, \rosat\ X-ray images are used
for comparison. The \xmm\ observation presented here allows for a refined
X-ray view and multiwavelength comparison of the cluster.

There is no dominant galaxy at the cluster center. Instead,
it contains several early-type galaxies of comparable brightnesses.
The gravitational mass distribution shows multiple peaks or filamentary 
structures around the X-ray centroid of the cluster \citep{2003ApJ...591..662D}. 
However, these peaks are well separated and are not well traced by 
the projected spatial distribution of galaxies, which
seems to be rather structured (e.g., see Fig.~13 in \citealt{2009A&A...495...15B}).
Such multiple-component structure is also seen in the velocity 
distribution of the cluster galaxies \citep{2009A&A...495...15B}. 

On even larger scales, one finds the so-called WSW galaxy group, 
$\sim$6$^{\prime}$ west (slightly to the south) of A959. This group 
is first detected via the gravitational lensing \citep{2003ApJ...591..662D} and
also in the \rosat\ image and galaxy concentration \citep{2009A&A...495...15B}.
Based on a diffuse X-ray spectrum extracted from the \xmm\ observation, we find that the temperature of the plasma in the group
is $T=1.7^{+2.3}_{-0.4}$ keV, indicating $M_{200}=0.7^{+1.7}_{-0.3} \times 10^{14} M_\odot$). The BCG (SDSS J101642.79+593223.3; z=0.285) of the group 
has a radio counterpart detected in both 
FIRST and NVSS. The redshift of the group is consistent with that of
A959. These two systems appear to be gravitationally bound, because 
their combined gravitational mass is greater than $2.6 \times 10^{14}M_{\odot}$,
required, according to the two-body model (\citealt{1982ApJ...257...23B}). Further away from A959 (beyond the \xmm\ field explored
here), there are two more clusters at comparable redshifts:
MaxBCG J153.93477+59.57870 at z$\sim$0.281 \citep{2007ApJ...660..239K}, 
$\sim$13$^{\prime}$ west, and MaxBCG J154.78308+59.76784 at z$\sim$0.284,
$\sim$18$^{\prime}$ NE (see Fig.~14 in \citealt{2009A&A...495...15B}). Therefore, 
A959 is apparently in a quite rich environment of the large-scale structure 
\citep{2009A&A...495...15B}.

{\sl A1084.} Its BCG (NVSS J104432-070407) is a radio source (F$_{1.4GHz}$= 31.6 mJy; \citealt{2007MNRAS.379..260M}) and hosts a significant star formation (SFR$\sim$0.4 $M_{\odot}$ yr$^{-1}$; \citealt{2010ApJ...715..881D}), which are consistent with the fact that a CC cluster (Fig.~\ref{fig:RP}) is more likely to host a BCG with star forming and/or radio emission (e.g., \citealt{2010ApJ...715..881D}, \citealt{2015A&A...581A..23K}).

{\sl A2813.} The BCG of this cluster does not show enhanced UV and mid-IR emission \citep{2012ApJS..199...23H}, which is consistent with its non-cool core (NCC) nature (Fig.~\ref{fig:RP}). There is a NW extension inferred from the X-ray image, probably due to an accretion of a filament \citep{2005A&A...442..827F}.

\subsection{Dynamical states of the clusters}
\label{ss:dyn}
Galaxy clusters are assembled via mergers of hierarchical systems, ranging
from the IGM to sub-clusters of galaxies at the intersections of cosmic web 
filaments. Therefore, depending on its recent merger history,
a cluster can be in quite different dynamical states.  

Because of the diversity in the recent merger histories of individual 
clusters, which is further complicated by their projection effects, 
their dynamical states are not straight-forward to determine. 
In Appendix~\ref{a;diag}, we present 
a review of various dynamical state diagnostics 
based on X-ray morphological properties of clusters.
The criteria for a relaxed cluster is the power ratio $P_3/P_0$ \citep{1995ApJ...452..522B}, centroid shift $w$ \citep{1993ApJ...413..492M},
surface brightness concentration $c_{SB}$ \citep{2008A&A...483...35S}, photon asymmetry $A_{\rm phot}$ \citep{2013ApJ...779..112N}, and
symmetry-peakiness-alignment SPA \citep{2015MNRAS.449..199M}.
The results of the applications of these diagnostics to our sample clusters are listed in Table~\ref{t:dynamic}.

We have also considered other cluster dynamical state tracers.
Clusters can be categorized into CC clusters and NCC clusters (e.g., \citealt{1984ApJ...276...38J}; \citealt{2006MNRAS.372.1496S}). A CC cluster tends to show a relaxed and symmetric morphology, whereas an NCC cluster likely exhibits a disturbed overall shape and substructure. We consider the presence of a central temperature 
drop as the evidence for a CC from Fig.~\ref{fig:RP}.

Another useful diagnostic for the dynamic state is the offset 
between the X-ray centroid of a cluster from its BCG.
It has been shown that the BCG tends to be found positionally at 
the gravitational center and kinematically near the rest frame of a cluster (e.g., \citealt{2016MNRAS.456.2566C}). Therefore, a large offset 
from the X-ray centroid of a cluster would indicate an ongoing
or recent merger activity (\citealt{1984ApJ...276...38J}; \citealt{2004ApJ...617..879L}). Previous studies show that an offset of $\sim$15 kpc may be 
used to distinguish a relaxed cluster from an un-relaxed one (e.g., offsets of $\leq$15 kpc as relaxed clusters in \citealt{2009MNRAS.398.1698S}; offsets center at $\sim$10/$\sim$50 kpc for relaxed/un-relaxed clusters in \citealt{2014MNRAS.439....2V}).
The X-ray/BCG offsets of our sample are listed in Table \ref{t:clusters}. 

We also incorporate information from observations at other wavelengths from \S~\ref{ss:multi}. Such information, though generally quite limited, may provide additional diagnostics of the cluster dynamical states. 

With the above diagnostics in consideration, we make general assessments
of the dynamics states of our sample clusters. We find that A1084 is a well 
relaxed cluster, which is the only one that passes all the criteria (morphological parameters; CC cluster; small X-ray/BCG offset; BCG hosts SF and AGN activities).
Other clusters all show multiple indications for major disturbance, although
each of them may meet at least one of the morphological 
parameters for a relaxed state. 
Their X-ray/BCG offsets are larger than 15 kpc, their temperature profiles do not show a strong center drop,
consistent with their disturbed dynamical states. 
This diversity in the apparent dynamical states reflects the complexity 
of cluster evolution and potentially the project effects, 
which limits the usefulness of any single diagnostic indicator.

\begin{table*}
\begin{threeparttable}
 \centering
  \caption{X-ray morphological parameters}
  \begin{tabular}{@{}lccccc@{}}
\hline \hline
Name & P$_{3}$/P$_{0}$ $\times$ 10$^{-7}$ & $w \times 10^{-2}$ & $c_{SB}$  & A$_{phot}$ & SPA \\
(1)  & (2) & (3) & (4)  & (5) & (6) \\
\hline
J0350 & $ 9.56^{+19.62}_{- 8.51}$ & $ 1.36^{+ 1.23}_{- 0.70}$ & $0.070^{+0.030}_{-0.026}$ & $ 0.09^{+ 0.02}_{- 0.06}$ & $0.813$ $-1.582$ $0.923$\\
A655 & $ 1.16^{+ 1.74}_{- 0.97}$ & $ 1.69^{+ 0.66}_{- 0.65}$ & $0.045^{+0.006}_{-0.007}$ & $ 0.06^{+ 0.01}_{- 0.02}$ & $0.533 \pm 0.171$ $-1.298 \pm 0.413$ $0.631 \pm 0.255$\\
A959 & $ 2.54^{+ 4.28}_{- 2.09}$ & $ 1.37^{+ 0.84}_{- 0.62}$ & $0.013^{+0.004}_{-0.004}$ & $ 0.48^{+ 0.24}_{- 0.43}$ & $0.876 \pm 0.152$ $-1.961 \pm 0.389$ $1.346 \pm 0.241$\\
A1084 & $ 0.27^{+ 0.16}_{- 0.10}$ & $ 0.31^{+ 0.02}_{- 0.02}$ & $0.164^{+0.003}_{-0.002}$ & $ 0.14^{+ 0.01}_{- 0.01}$ & $1.158 \pm 0.029$ $-0.814 \pm 0.008$ $1.356 \pm 0.034$\\
A2813 & $ 0.41^{+ 0.75}_{- 0.28}$ & $ 1.33^{+ 0.14}_{- 0.15}$ & $0.040^{+0.003}_{-0.003}$ & $ 0.05^{+ 0.02}_{- 0.03}$ & $1.112 \pm 0.081$ $-1.040 \pm 0.007$ $1.193 \pm 0.058$\\
\hline
\end{tabular}
\begin{tablenotes}
\item {\sl Note.} (1) cluster name. (2) Power ratio. (3) Centroid shift. (4) Surface brightness concentration. (5) Photon asymmetry. (6) Symmetry-Peakiness-Alignment. More details in Appendix~\ref{a;diag}.
\end{tablenotes}
\label{t:dynamic}
\end{threeparttable}
\end{table*}

\subsection{Baryon contents}
Here we first estimate for each cluster the masses of the three main cluster components: dark matter, ICM, and stars. We will then compare the total baryon budgets with the values expected from the standard cosmology. 

\subsubsection{Total gravitational mass}
\label{ss:total mass}

We use the $M-T$ relation (\citealt{2009ApJ...693.1142S} for \cha; \citealt{2007A&A...474L..37A} for \xmm) to estimate the total cluster mass $M_{500}$ in $r_{500}$, as the X-ray temperature is a quite robust mass proxy. However, this X-ray hydrostatic mass could be underestimated because of additional non-thermal pressure from e.g., gas bulk motion and turbulence (e.g., \citealt{2007ApJ...668....1N}; \citealt{2008MNRAS.384.1567M}; \citealt{2009ApJ...705.1129L}).
Assuming a typical 15\% hydrostatic bias (X-ray hydrostatic mass is 15\% lower than the true cluster mass) in total cluster mass induces to 21\% in $M_{500}$ \citep{2014MNRAS.438...78R}.
Moreover, most clusters in our sample are unrelaxed clusters, thus the hydrostatic bias could be even higher.

We note there are some cross-calibration issues between \cha\ and \xmm, especially the temperature from \cha\ are systematically higher than \xmm\ (e.g., \citealt{2015A&A...575A..30S}). We use the temperature relation of ACIS and combined \xmm\ in the full band in \cite{2015A&A...575A..30S} to rescale the temperature of \cha\ to
\xmm, and then re-estimate the $M_{500}$ based on the \xmm\ $M-T$ relation, the mass difference is within 30\%. However, our final results remain unchanged when including the above bias.
\subsubsection{Hot ICM mass}

\label{ss:mgas}
For the hot ICM, we calculate the central de-projected proton (or hydrogen) density distribution from our $\beta-$model fit to the radial intensity profile (e.g., \citealt{1988xrec.book.....S}), under assumptions of the spherical symmetry and isothermal temperature distribution (the cluster emissivity function in soft 0.5-2 keV band is sensitive to density but almost insensitive to temperature at $T > 2$ keV),
\begin{equation}
n_{H}=n_0 \Big(1+\frac{r^2}{r_{c}^{2}}\Big)^{-\frac{3}{2}{\beta}}.
\end{equation}
We adopt the equation (10) of \cite{2016MNRAS.459..366G} to estimate $n_0$:
\begin{equation}
n_0=\frac{180}{\pi}\sqrt{\frac{10^{14}4\sqrt{\pi}I_0\Gamma(3\beta)}{(\frac{n_e}{n_H})\frac{CR}{\eta}r_c\Gamma(3\beta-1/2)}}\ (XMM\text{-}Newton\ {\rm data}),
\end{equation}
\begin{equation}
n_0=\frac{3600 \times 180}{\pi}\sqrt{\frac{10^{14}4\sqrt{\pi}I_0\Gamma(3\beta)}{(\frac{n_e}{n_H})\frac{f_x}{\eta}r_c\Gamma(3\beta-1/2)}}\ (Chandra\ {\rm data}).
\end{equation}
These equations use the parameters from the $\beta$-model fit ($I_0$, $\beta$, $r_c$ in a unit of cm) of the exposure-corrected intensity image, and the spectral dependent conversion from the emission measure to the count rate ($\frac{CR}{\eta}$ or $\frac{f_x}{\eta}$) with the on-axis response files of each observation. This is different from the method used in such works as \cite{1999ApJ...517..627M} and \cite{2004A&A...417...13E}, which integrate the temperature-dependent electron density along the line of sight within a somewhat artificial cluster ``boundary", where the X-ray emission is insignificant. In contrast, our analytical formula accounts for all the X-ray emission along the line of sight. The $n_0$ and the hot ICM mass ($M_{gas}$)
within $r_{500}$ are listed in Table~\ref{t:clusters}.

The systematic error of $M_{gas}$ is dominated by the uncertainty from integration gas distribution in the outer radius, which is closely related to the $\beta$ value. The $\beta$ varies with cluster mass (e.g., \citealt{2009ApJ...693.1142S}) and radius (e.g., \citealt{2015MNRAS.450.2261M}). We assume a typical 10\% uncertainty in $\beta$ to estimate the error in $M_{gas}$. Moreover, the gas mass could be overestimated due to bias of gas clumping, though it should be small within $r_{500}$.

For future comparison, Table~\ref{t:clusters} also lists the estimation of the hot ICM proton column density at the projected distances of the QSO for each cluster as
\begin{equation}
N_p=\sqrt{\pi}n_0 r_c\frac{\Gamma(3\beta/2-1/2)}{\Gamma(3\beta/2)}(1+x^{2})^{1/2-3\beta/2}\ \ \  (\beta>1/3).
\end{equation}

\subsubsection{ Warm-hot ICM}

\label{ss:mgas}
\begin{figure}
     \begin{center}
\includegraphics[width=0.5\textwidth,keepaspectratio=true,clip=true]{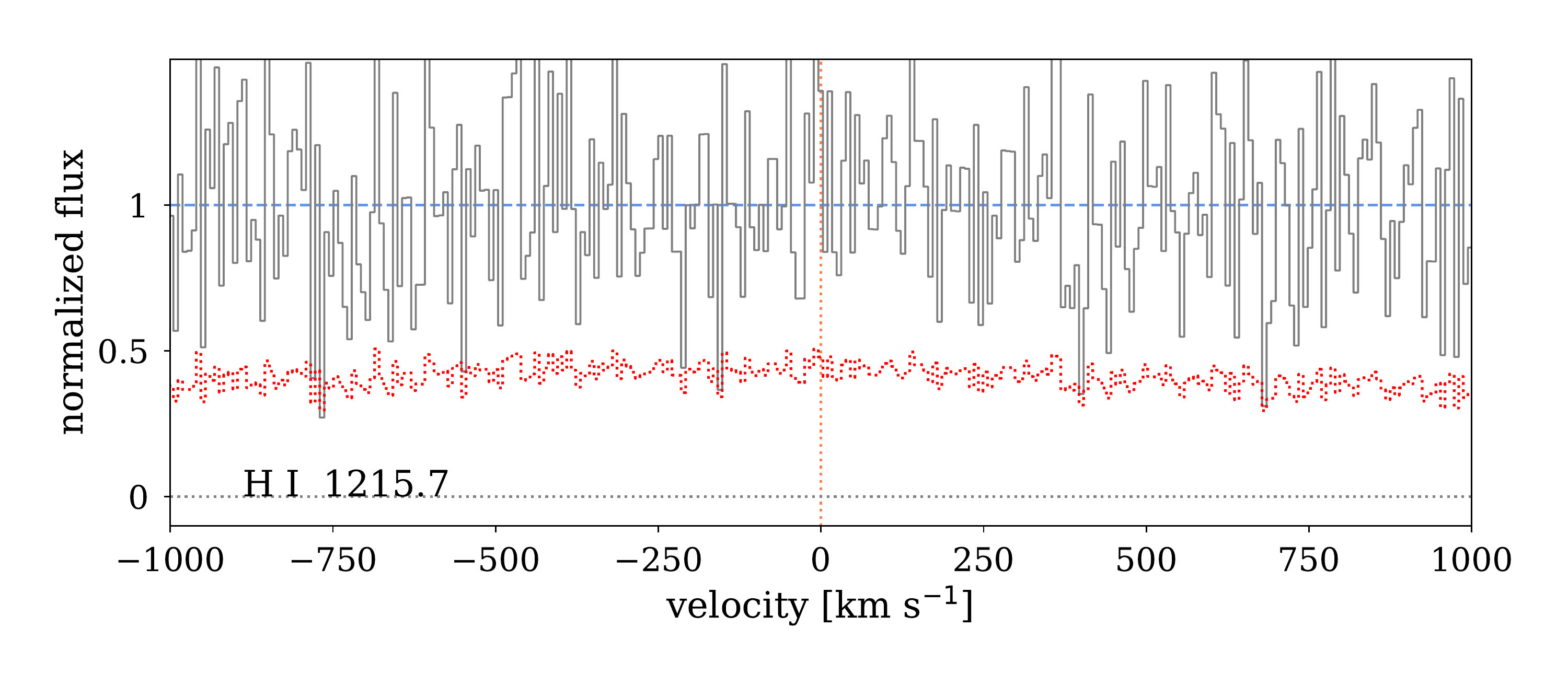}
    \end{center}
    \caption{Continuum-normalized spectrum of SBS 1013+596 in the region where H~\textsc{I} Ly$\alpha$ would fall at the redshift of A959, to which the velocity scale is relative.  The error vector is shown red.  While the spectrum is quite noisy, we do not detect any statistically significant Ly$\alpha$ features within $\pm$ 1000 km s$^{-1}$ of the cluster redshift.}%
   \label{f:sbs1013spec}
\end{figure}

The larger goal of the work presented herein and in our previous publications \citep{2014MNRAS.439.1796W,2016MNRAS.459..366G,2018MNRAS.475.2067B} is to leverage diagnostics of both the hot ($T > 10^6$ K) and warm-hot ($T = 10^{5-6}$ K) gas afforded by X-ray imaging/spectroscopy and UV absorption lines, respectively, to characterize the ICM in its baryon budget and impact on the circumgalactic media of cluster galaxies.  We now present an analysis of the existing HST/COS spectrum of SBS 1013+596, whose sightline probes the outskirts of A959 at an impact parameter of 2.68 Mpc, or 2.58 $r_{500}$.  Our analysis follows closely that of \citet[][Section 4.2]{2018MNRAS.475.2067B}.  The SBS 1013+596 spectrum was obtained in HST programme GO 12593 (PI: Nestor) using the COS G160M grating, covering the wavelength range of 1410-1774 \AA\ with an average S/N of 2.4 in the region of interest (near the observed wavelength of the H~\textsc{I} Ly$\alpha$ line at $z=0.288$.  Due to the wavelength coverage of the G160M grating, H~\textsc{I} Ly$\beta$ and the O~\textsc{vi} are not accessible. Fig.~\ref{f:sbs1013spec} shows the QSO spectral region within $\pm$ 1000 km s$^{-1}$ of the Ly$\alpha$ observed wavelength at the redshift of A959.

\begin{figure}
     \begin{center}
\includegraphics[width=0.5\textwidth,keepaspectratio=true,clip=true]{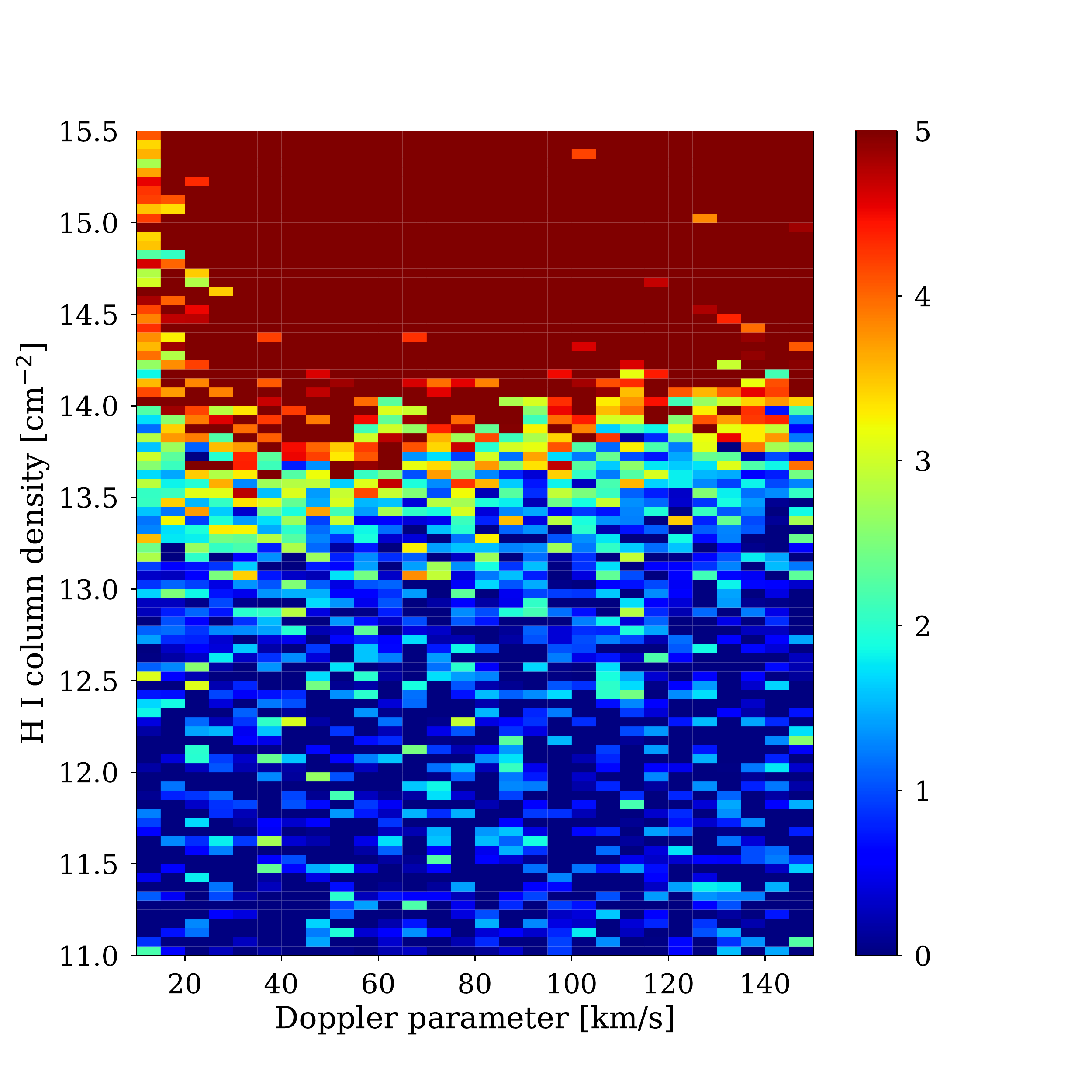}
    \end{center}
    \caption{Detection significance (see colorbar on the right) of synthetic H~\textsc{i} Ly$\alpha$ features at the redshift of A959 as a function of H~\textsc{i} column density and Doppler $b$ value.  Assuming the lines are thermally broadened, one may associate these $b$ values to gas temperature.  Note that while lines become more difficult to detect with increasing $b$, with the column density limits increasing by 0.5 dex over the range of Doppler parameter, the poor S/N of this archival spectrum hardly enable detecting lines of N(H~\textsc{i}) $< 13.5$ cm$^{-2}$ for even narrow Ly$\alpha$ lines.}%
   \label{f:sbs1013detsig}
\end{figure}

We report no statistically significant detections of H~\textsc{i} Ly$\alpha$ near the redshift of A959, possibly due in part to the low S/N of the archival SBS 1013+596 spectrum. However, we proceed with the analysis detailed by \citet{2018MNRAS.475.2067B} to place constraints on the physical conditions and total column density of the 10$^{5-6}$ K gas.  Briefly, the analysis premise is as follows: assuming thermal broadening of any potential Ly$\alpha$ lines, we generate numerous realizations of Voigt profiles on a grid of Doppler $b$-parameters and column densities (N(H~\textsc{i})).  We then convolve these synthetic profiles with the instrumental line spread function and inject them into a flat spectrum with S/N commensurate with that of the actual data.  Finally, measuring the equivalent widths and associated uncertainties of these profile realizations enable us to quantify their detection significance.  Fig.~\ref{f:sbs1013detsig} shows the matrix of detection significance for the range of N(H~\textsc{i}) and $b$, where the colors indicate the detection significance according to the values shown on the colorbar.  We encourage the reader to contrast Fig.~\ref{f:sbs1013detsig} to its counterpart in \citet[][Fig.~10]{2018MNRAS.475.2067B}, as the quantitative difference in data quality between those targeted, S/N $\sim 18$ spectra and the archival spectrum analysed here is clearly seen.  The \citet{2018MNRAS.475.2067B} data enabled sensitivity to column densities approximately an order of magnitude lower than in the present work (N(H~\textsc{i})$\sim 10^{12.5} {\rm cm}^{-2}$ vs. $\sim 10^{13.5} {\rm cm}^{-2}$ for $b=40$ km s$^{-1}$ lines).  Given that the $b \sim 70$-km s$^{-1}$ broad Ly$\alpha$ lines associated with A1926 reported by \citet{2018MNRAS.475.2067B} had N(H~\textsc{i})$< 10^{13.7}~{\rm cm}^{-2}$, only the strongest component would be possibly (but unlikely) detected, and the remaining two N(H~\textsc{i})$\sim 10^{13.1} {\rm cm}^{-2}$ lines would surely be lost in the noise.  Recall that two out of three of the \citet{2018MNRAS.475.2067B} sightlines were targeted with appropriate S/N to robustly detect such features and probe X-ray bright galaxy clusters at interesting impact parameters; the third is among the highest S/N QSO spectra ever recorded by HST/COS.  We emphasize that chance alignments alone between massive clusters and QSO sightlines that have been observed with sufficient data quality in the HST archive are simply insufficient to build samples and conduct the analysis necessary to study the warm-hot gas contribution to cluster baryon budgets.  

\begin{figure}
     \begin{center}
\includegraphics[width=0.5\textwidth,keepaspectratio=true,clip=true]{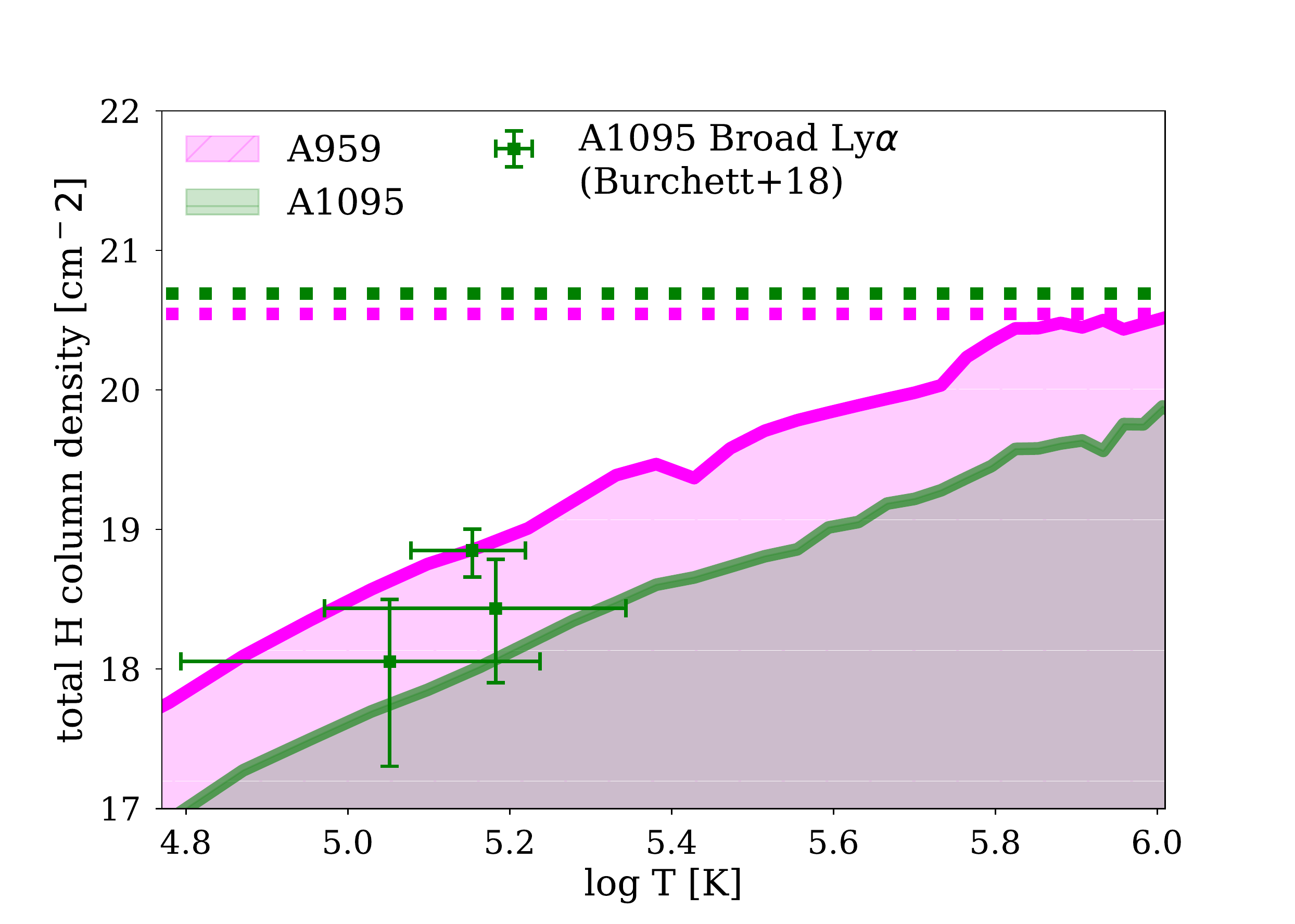}
    \end{center}
    \caption{Limits on the total H column density from gas in the warm-hot gas phase as a function of temperature under thermal line broadening (solid lines and shaded regions) relative to that from X-ray measurements of the $>10^6$ K phase (dotted horizontal lines).  Measurements corresponding to A959/SBS 1013+596 are colored purple, and we show the limits and measurements derived for A1095 \citep{2018MNRAS.475.2067B} in green for comparison. In this representation, the QSO spectra enable detecting N(H) values above the solid lines, and N(H) values lying in the shaded regions would be obscured by the noise in the data.  Note that the three previously detected components associated with A1095 would be undetectable  in A959 given the S/N of the existing spectrum.  In A959, the amount of total H column hidden in 10$^{5.8}$ K gas could equal that in the hot X-ray traced material.}%
   \label{f:wahogalimits}
\end{figure}

Finally, we convert the detection statistics as functions of $b$ value and column density to total H column density (N(H)) to compare with the constraints from X-rays.  As in \citet{2018MNRAS.475.2067B}, we use the \citet{2013MNRAS.434.1062} ionization models to obtain the ionization fractions H$^{o}$/H as a function of temperature (obtained from the Doppler $b$ values).  The N(H) is then the measured N(H~\textsc{i}) divided by the ion fraction.  Fig.~\ref{f:wahogalimits} shows the results, comparing the limits we are able to place on A959 with the constraints \citet{2018MNRAS.475.2067B} placed on A1095.  Here, the solid line shows the 3-$\sigma$ detection threshold as a function of temperature, and the shaded regions below these lines represent the amount of the N(H) that could be `hidden' in the noise of the data.  The markers with error bars represent the individual components of broad Ly$\alpha$ associated with A1095.  Clearly, these components would be missed in spectra with similar S/N to that of SBS 1013+596 analyzed here.  We suggest that these data serve as a useful guide to establish observational strategies in pursuing these simultaneous hot/warm-hot constraints with a targeted sample of the cluster/QSO sightline pairs.

\subsubsection{Baryon fraction of the clusters}

Fig.~\ref{f:fbaryon} shows the total baryon fraction over the total gravitational mass of the hot ICM and stars. 
The stellar mass is estimated with the scaling relation of $M_{\star,3D}=3.2\times10^{-2}(M_{500}/10^{14}\ M_{\odot})^{0.52}$ from \cite{2013ApJ...778...14G}, whose estimate includes the additional contribution from intracluster stars (ICS) and correction for the completeness from fainter galaxies as well as the projection from galaxies outside $r_{500}$ in the line of sight direction. The stellar fractions of the individual cluster are also presented in Fig.~\ref{f:fbaryon}. {Though there are some systematic biases when determining stellar mass, e.g., initial mass function (IMF) variation (radial and galaxy dependent); ICS contribution; uncertainty in mass-to-light ratio depending on different models and methods; uncertainty in the faint end slope of the galaxy luminosity function; and deprojection of galaxy distribution from 2D to 3D (e.g., \citealt{2013ApJ...778...14G}; \citealt{2014MNRAS.437.1362B}; \citealt{2018AstL...44....8K}). The scatter of the stellar mass is at a level of 30\% (\citealt{2004ApJ...617..879L}; \citealt{2018AstL...44....8K}). Thus the biases in stellar mass estimation have a minor impact to our total baryon fraction, because the dominant baryon component is in hot gas for our cluster sample.

For comparison, Fig.~\ref{f:fbaryon} further includes results from our previous studies of four clusters \citep{2016MNRAS.459..366G}, which are all disturbed clusters. We account for the 21\% hydrostatic bias in $M_{500}$ as discussed in \S~\ref{ss:total mass}. As the $M_{500}$ increasing, the $r_{500}$ for estimating $M_{gas}$ and $M_{star}$ also increases. However, the $f_{gas}$, which dominates the baryon budget, only increase 2.6\% with $f_{gas} \sim M_{500}^{0.135}$ \citep{2009ApJ...693.1142S}, while $f_{star}$ decreases with \cite{2013ApJ...778...14G} relation. All the clusters show baryon deficiency compared with the cosmological fraction determined from the Wilkinson Microwave Anisotropy Probe (WMAP) 9-year data ($\sim$17\%; \citealt{2013ApJS..208...19H}).

A portion of this missing baryon matter may be ejected out of the central $r_{500}$ regions due to energetic processes such as AGN feedback (e.g., \citealt{2011MNRAS.412.1965M}), while some fraction may reside in other gas phases, such as the warm and warm-hot ICM. As demonstrated in the previous subsection and by \citet{2018MNRAS.475.2067B}, high-quality, targeted QSO absorption line observations will help to characterize or constrain these missing baryons.

\begin{figure}
     \begin{center}
\includegraphics[width=0.45\textwidth,keepaspectratio=true,clip=true]{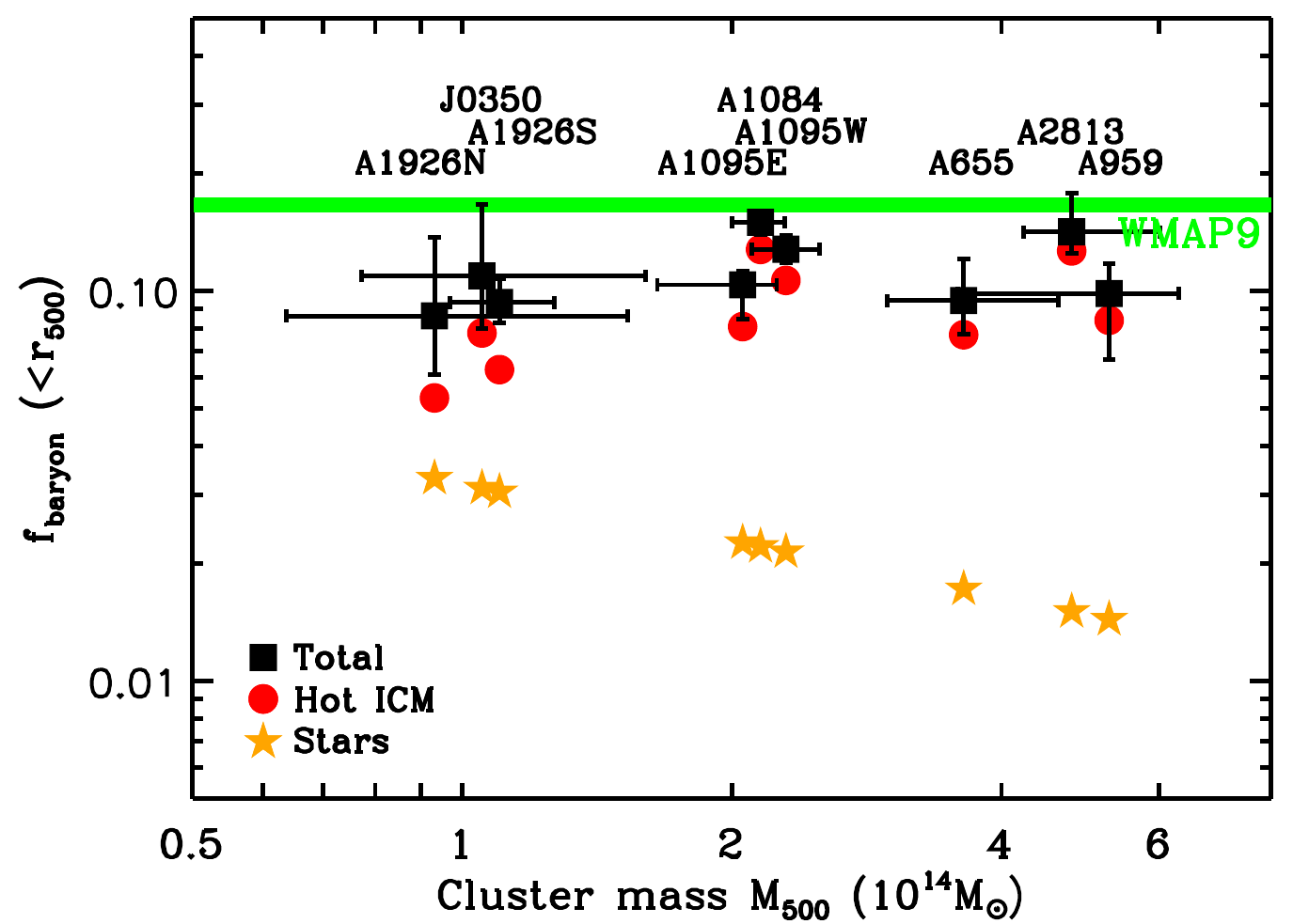}
    \end{center}
    \caption{
Baryon fraction within $r_{500}$ as a function of $M_{500}$: dots are the hot ICM fractions; stars represent the stellar mass fractions estimated based on the scaling relation of \protect\cite{2013ApJ...778...14G}, which includes the additional contribution from ICS; boxes are the total baryon fractions of hot ICM plus stars. The green band marks the uncertain range of the WMAP 9-year baryon fraction \protect\citep{2013ApJS..208...19H}. The above labels are names of corresponding clusters. We also include four clusters in \protect\cite{2016MNRAS.459..366G}.}%
   \label{f:fbaryon}
\end{figure}

\section{Summary}
\label{s:sum}

We have studied a sample of X-ray-selected galaxy clusters that are paired with UV-bright background QSOs. While this study is based primarily on \cha\ and \xmm\ observations,　we have also incorporated the single existing archival HST/COS spectrum for the QSO sightlines, as well as radio and optical observations for all, in the analysis of the baryon content and dynamical state of the clusters. 
Our major results and conclusions are as follows:

\begin{itemize}

\item We have characterized the X-ray morphological and spectral properties of
the clusters. The radial profile of the diffuse X-ray emission can be 
reasonably well fitted with a $\beta$-model out to $r_{500}$ except for A655, due to the 
shallow exposure. A 
single-temperature plasma describes well the average spectra of the clusters.

\item Most of the clusters are dynamically disturbed, judging from their quantitative X-ray morphological parameters, temperature profiles, X-ray/BCG offsets, diffuse radio features, and BCG activities. These disturbances may lead to underestimation of the total gravitational masses of the clusters. 

\item The HST/COS spectrum allows us to place only weak constraints on the warm-hot gas content of A959. This analysis demonstrates the need to obtain high S/N spectra of these QSOs in a focused observing programme.

\item All the clusters show an evidence for missing baryons considering the hydrostatic bias.
They may be located outside of the cluster regions explored here. Alternatively,
they may reside in the warm and warm-hot ICM, which are yet to be discovered.

\end{itemize}
With these in consideration, targeted UV absorption line spectroscopy of the
background QSOs will help us to characterize how the heating/cooling may depend on the dynamical state and mass of the cluster, as well as the impact 
distances, advancing our understanding of the structure formation 
and baryon evolution in the Universe.

\section*{ACKNOWLEDGEMENTS}
C.G. acknowledges support from the National Natural Science Foundation of China (No. 11703090).
Z.L. acknowledges support from the Recruitment Program of Global Youth Experts.
Q.G. is supported by the National Key Research and Development Program of China (No. 2017YFA
0402703), and by the National Natural Science Foundation of China (No. 11733002).
L.J. acknowledges support from the Joint Funds of the National Natural Science Foundation of China 
under grant U1531248.

\appendix
\section{Dynamical state diagnostics 
of galaxy clusters: X-ray morphological parameters}
\label{a;diag}
Here we provide a systematical review of various diagnostics based on the X-ray morphological parameters. The used quantitative morphological indicators generally fall into two categories: (1) the bulk asymmetry, e.g., the power ratio \citep{1995ApJ...452..522B}, the centroid shifts \citep{1993ApJ...413..492M}, the photon distribution asymmetry (A$_{\rm phot}$, \citealt{2013ApJ...779..112N}); (2) the presence of a CC, e.g., the surface brightness concentration parameter \citep{2008A&A...483...35S}. Apparently, combining the above two categories can provide a better separation of clusters at different dynamical states. An example is the symmetry-peakiness-alignment (SPA) criterion for relaxation \citep{2015MNRAS.449..199M}. Here, we briefly describe these morphological parameters and their relations.

The power ratio method (see details in \citealt{1995ApJ...452..522B}; \citealt{2005ApJ...624..606J}) is motivated by identifying the X-ray surface brightness as a representation of the projected mass distribution of a cluster, which aims to parametrize the substructure in the ICM and to relate it to the dynamical state of a cluster. The power ratio $P_3/P_0$ has been found to be sensitive to asymmetries in the aperture radius ($r_{500}$ in our case) and provides a useful measure of the dynamical state of a cluster. We use the cut of \cite{2010A&A...514A..32B} to separate relaxed cluster ($P_3/P_0 \leq 1.5\times10^{-7}$) and disturbed cluster ($P_3/P_0 > 1.5\times10^{-7}$).

The centroid shift parameter $w$ (\citealt{1993ApJ...413..492M}; \citealt{2006MNRAS.373..881P}) measures the centroid variations in different aperture sizes and is defined as the standard deviation of the different center shifts (in units of $r_{500}$).
We also use the cut of \cite{2010A&A...514A..32B} to separate relaxed cluster ($w \leq 0.01$) and disturbed cluster ($w > 0.01$).

The surface brightness concentration parameter $c_{SB}$ \citep{2008A&A...483...35S} measures the ratio of the peak
over the ambient surface brightness and is very effective to distinguish CC from NCC clusters. 
We use three  categories of \citep{2008A&A...483...35S}: non-CC ($c_{SB} < 0.075$), moderate ($0.075 < c_{SB} < 0.155$), and pronounced ($c_{SB} > 0.155$) CC.

The photon distribution asymmetry parameter $A_{\rm phot}$ \citep{2013ApJ...779..112N} measures the uniformity of the angular X-ray photon distribution in radial annuli. The parameter quantifies the deviation from the idealized axisymmetric case.
We use the threshold values of \cite{2013ApJ...779..112N} to separate low ($A_{\rm phot}<0.15$), medium ($0.15<A_{\rm phot}<0.6$), and strong asymmetry ($A_{\rm phot}>0.6$) clusters. As both $A_{\rm phot}$ and $w$ measure the bulk asymmetry on intermediate scales (in $r_{500}$), thus they are correlated strongly (Fig.~8 in \citealt{2013ApJ...779..112N}). Furthermore, these asymmetry indicators, combined with a CC indicator such as concentration $c_{SB}$, show a better separation of clusters at different states
of dynamical equilibrium, e.g., the asymmetry-concentration diagram (Fig.~7 in \citealt{2013ApJ...779..112N}).

The symmetry (s)-peakiness (p)-alignment (a) (\citealt{2015MNRAS.449..199M}) method classifies the cluster dynamical state based on both asymmetry and CC indicators. 
The $p$ contains the information of the flux ratio in small and large apertures, and is similar to the definition of $c_{SB}$, thus they correlate strongly with each other (Fig.~9 in \citealt{2015MNRAS.449..199M}). 
The $s$ and $a$ have similar definitions with centroid shift, and they all measure bulk asymmetry, thus they correlate strongly with each other (Figs.~7-8 in \citealt{2015MNRAS.449..199M}). The combination of asymmetry indicators such as $s$ and $a$ with a CC indicator such as $p$ separates the relaxed cluster from un-relaxed cluster well (Figs~.8,12 in \citealt{2015MNRAS.449..199M}). We use the cuts of $s > 0.87$, $p > -0.82$, and $a > 1.00$ \citep{2015MNRAS.449..199M} as the criterion for a relaxed cluster.

\begin{figure}
 	\begin{center}
\includegraphics[width=0.45\textwidth,keepaspectratio=true,clip=true]{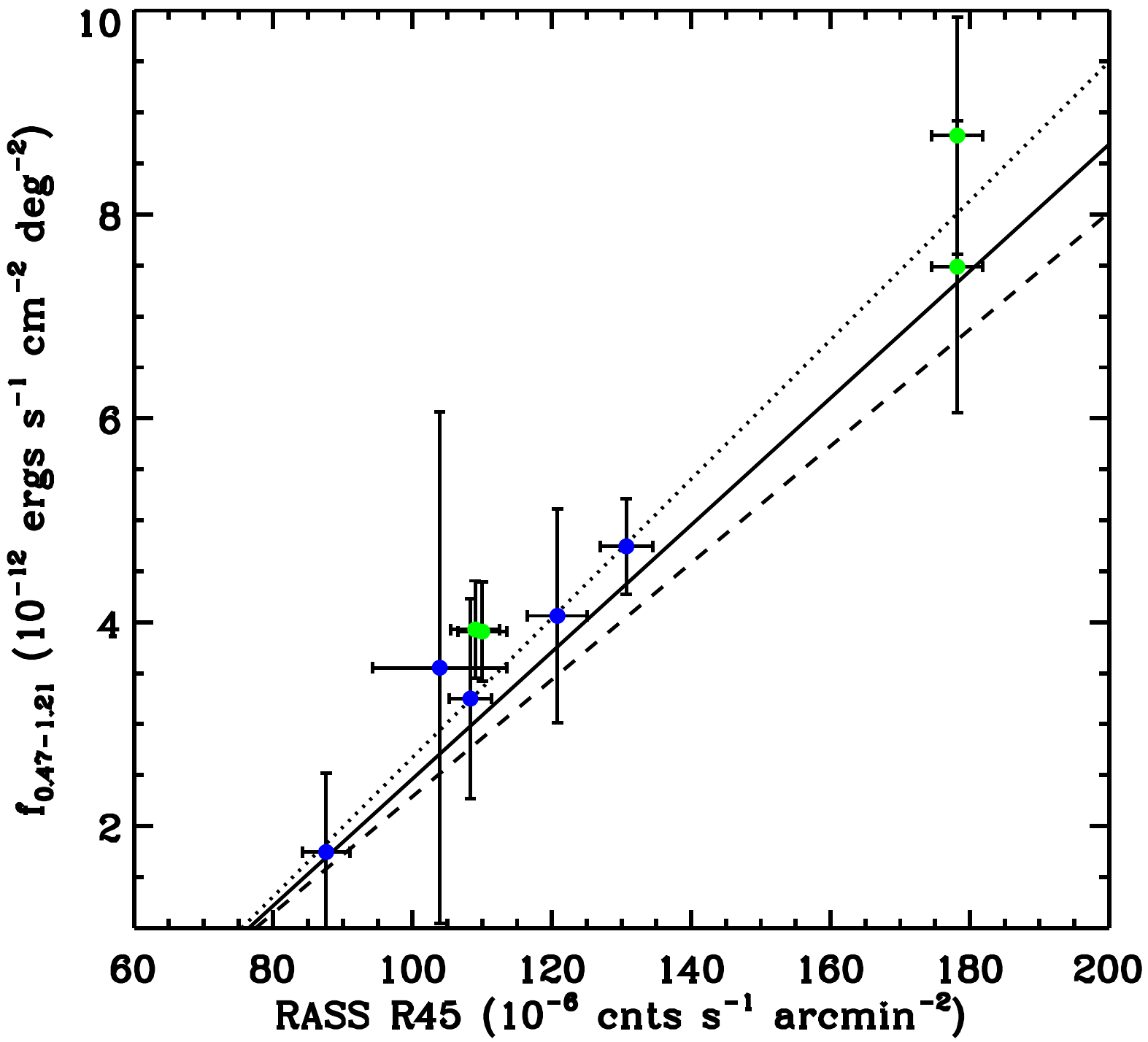}
 	\end{center}
	\caption{
0.47-1.21 keV on-cluster sky flux vs. off-cluster RASS R45 (0.47-1.21 keV) flux. Blue dots represent five clusters in the present sample, while green dots are four clusters from \citep{2016MNRAS.459..366G}. Error bars are at the 90\% confidence level. The three lines are from \citep{2009ApJ...693.1142S}, with the expected conversions between two fluxes, assuming two thermal components for the soft sky background (all with $T_{cool}$ = 0.1 keV). The solid line is for $T_{hot}$ = 0.25 keV and NORM$_{hot}$
/ NORM$_{cool}$ = 0.5. The dotted line is for $T_{hot}$ = 0.2 keV and NORM$_{hot}$
/ NORM$_{cool}$ = 0.5. The dashed line is for $T_{hot}$ = 0.3 keV and NORM$_{hot}$ / NORM$_{cool}$ = 1.}
	 \label{fig:bkg}
\end{figure}

\end{document}